\documentclass[english,11pt,aps,prd,a4paper,preprintnumbers, floatfix,nofootinbib,showpacs,superscriptaddress,  notitlepage]{revtex4-1} 
 \pdfoutput=1
\usepackage[usenames,dvipsnames]{color}  %%%%%%% usenames,dvipsnames required to get BrickRed
\usepackage{graphicx}
\usepackage{setspace}
\usepackage{upgreek}
\usepackage{braket}
\usepackage{caption}
\usepackage{stackengine}
\usepackage{subcaption}
\usepackage{amsmath}
\usepackage{amssymb}
\usepackage[colorlinks=true,citecolor=darkred,urlcolor=darkred, pdfborder={0 0 0}]{hyperref}
\usepackage[normalem]{ulem}
\usepackage{xcolor}
\usepackage{graphicx}
\makeatletter
\def\p@subsection{}
\makeatother
\usepackage{xcolor}
\usepackage{float}

\usepackage{soul}

\definecolor{darkred}{rgb}{0.6,0,0}

\definecolor{linkcolor}{rgb}{0,0,0.5}

\def\gsim{\raise0.3ex\hbox{$\;>$\kern-0.75em\raise-1.1ex\hbox{$\sim\;$}}}
\def\lsim{\raise0.3ex\hbox{$\;<$\kern-0.75em\raise-1.1ex\hbox{$\sim\;$}}}

\def\beqn#1{\begin{equation}\label{#1}}
\def\eeqn{\end{equation}}

\def\beqa#1{\begin{eqnarray}\label{#1}}
\def\eeqa{\end{eqnarray}}

\usepackage{caption}
\usepackage{subcaption}
\usepackage{adjustbox}
\def\Z2{$\mathcal{Z_2}$}

\usepackage{bbold}
\usepackage{ragged2e}
\newcommand {\ignore}[1]{}

\newcommand{\sm}{{Standard Model }}

\def\cevns{CE$\nu$NS }

\usepackage{mathrsfs}
\makeatletter
\newcommand\footnoteref[1]{\protected@xdef\@thefnmark{\ref{#1}}\@footnotemark}
\makeatother
\def\321{$\mathrm{SU(3) \otimes SU(2) \otimes U(1)}$ }

\definecolor{vdrgreen}{rgb}{0.0, 0.7, 0.0}

\newcommand{\AddrIISERB}{Department of Physics, Indian Institute of Science Education and Research - Bhopal, \\ 
Bhopal Bypass Road, Bhauri, Bhopal 462066, India}

\newcommand{\AddrAthens}{%
Department of Physics, National and Kapodistrian University of Athens, Zografou Campus GR-15772 Athens, Greece}

\newcommand{\AddrAHEP}{%
  AHEP Group, Institut de F\'{i}sica Corpuscular --
  CSIC/Universitat de Val\`{e}ncia \\ % Parc Cient\'ific de Paterna.\\
  C/Catedr\'atico Jos\'e Beltr\'an, 2 E-46980 Paterna, Spain}

\begin{document}

\title{\textcolor{BrickRed}{Physics implications of recent Dresden-II reactor data}}
\author{Anirban Majumdar}\email{anirban19@iiserb.ac.in}
\affiliation{\AddrIISERB}
\author{Dimitrios K. Papoulias}\email{dkpapoulias@phys.uoa.gr}
\affiliation{\AddrAthens}
\author{Rahul Srivastava}\email{rahul@iiserb.ac.in}
\affiliation{\AddrIISERB}
\author{Jos\'e W. F. Valle}\email{valle@ific.uv.es}
\affiliation{\AddrAHEP}

\begin{abstract}

  Prompted by the recent Dresden-II reactor data we examine its implications for the determination of the weak mixing angle, paying attention to the effect of the quenching function.
  We also determine the resulting constraints on the unitarity of the neutrino mixing matrix, as well as on the most general type of nonstandard neutral-current neutrino interactions.

\end{abstract}
\maketitle

\section{Introduction}

The neutral-current observation of neutrino events opens new windows that may reveal novel aspects of neutrino physics.
Coherent elastic neutrino nucleus scattering (CE$\nu$NS) is a low-energy process, proposed by Freedman~\cite{Freedman:1973yd}, where the entire nucleus is scattered off elastically by the neutrinos.
Despite the large predicted \cevns cross section, its detection faces experimental challenges mainly because of the very tiny nuclear recoil signals produced in the aftermath of a \cevns event.
\cevns was first observed at the Oak Ridge National Laboratory by the COHERENT Collaboration which exploited the $\pi$-DAR (pion-decay-at-rest) neutrino beam coming from the Spallation Neutron Source facility (SNS) using a CsI detector~\cite{COHERENT:2017ipa, COHERENT:2021xmm} and subsequently using a liquid argon detector~\cite{COHERENT:2020iec}.
Nuclear power plants constitute another important experimental probe for CE$\nu$NS studies, with several experiments currently underway such as CONNIE~\cite{CONNIE:2021ggh}, CONUS~\cite{CONUS:2020skt, CONUS:2021dwh} and $\nu$GEN~\cite{nuGeN:2022bmg} or in preparation (e.g. MINER~\cite{MINER:2016igy}, RICOCHET~\cite{Billard:2016giu}, $\nu$-cleus~\cite{Strauss:2017cuu}, TEXONO~\cite{Wong:2015kgl}, $\nu$IOLETA~\cite{Fernandez-Moroni:2020yyl} and Scintillating Bubble Chamber (SBC)~\cite{SBC:2021yal}).
These facilities have the advantage of producing an extremely intense beam of low-energy antineutrinos ($E_{\nu}<10$ MeV). 
Thus, because of the very low energy-momentum transfer involved, the expected \cevns signal at reactors does not suffer from nuclear physics uncertainties, in contrast to $\pi$-DAR based
experiments~\cite{Baxter:2019mcx, CCM:2021yzc}. 

On the other hand, reactor experiments are subject to large neutron backgrounds, which together with the large quenching factor (QF) uncertainties are the main limiting factors for \cevns detection at reactor facilities. 

Very recently, the Dresden-II Collaboration~\cite{Colaresi:2021kus, Colaresi:2022obx} has reported a suggestive piece of evidence pointing to the first ever observation of \cevns with reactor antineutrinos, using a 3~kg germanium detector, namely NCC-1701 (Neutrino Coherent Coupling-$1701$).
The experiment is located $10.39$ meters away from the Dresden-II boiling water reactor and collected beam-on data for a total time of $96.4$ days, operating with a very low threshold of $0.2\text{ keV}_{ee}$. 
The new Dresden-II data have prompted phenomenological analyses that resulted in complementary constraints on various parameters within and beyond the Standard Model (SM).
The relevant works focused on the determination of the weak mixing angle~\cite{AristizabalSierra:2022axl, Khan:2022jnd, AtzoriCorona:2022qrf} as well as on popular SM extensions such as those involving electromagnetic neutrino interactions~\cite{AristizabalSierra:2022axl, Khan:2022jnd, AtzoriCorona:2022qrf} and light mediators ~\cite{AristizabalSierra:2022axl, Liao:2022hno, Coloma:2022avw}. 
Nonstandard interactions (NSIs) were also explored in Refs.~\cite{Coloma:2022avw, Denton:2022nol}.
They are a sideshow of the physics associated with low-scale neutrino mass generation~\cite{Boucenna:2014zba}.
Indeed, neutrino oscillation experiments have provided robust evidence for massive neutrinos, prompting many efforts for underpinning the origin of their masses.
A remarkable example is the low-scale seesaw mechanism, such as the inverse~\cite{Mohapatra:1986bd,Gonzalez-Garcia:1988okv} and the linear seesaw~\cite{Akhmedov:1995ip,Akhmedov:1995vm,Malinsky:2005bi}. 
% %
 In both of these, we have massless neutrinos in the limit of lepton-number conservation. They lead to small, symmetry-protected neutrino masses, mediated by heavy quasi-Dirac neutrino exchange.
 These schemes can be tested directly at high energies~\cite{Dittmar:1989yg,Gonzalez-Garcia:1990sbd,AguilarSaavedra:2012fu,Das:2012ii,Deppisch:2013cya},
 a possibility taken up by the ATLAS and CMS collaborations at the LHC~\cite{ATLAS:2019kpx,CMS:2018iaf,CMS:2022nty},
 or in future proposals---see, e.g., Refs.~\cite{Drewes:2019fou,Abdullahi:2022jlv}.
 Here we focus on their associated signals at low-energy precision studies of neutrino properties.
 An example is unitarity violation in neutrino mixing, as expected in a low-scale seesaw mechanism.\\[-.5cm]

 Building upon previous work~\cite{AristizabalSierra:2022axl}, in this paper we revisit the impact of Dresden-II data on the determination of the weak mixing angle.
In so doing, we adopt an improved analysis taking into account the full Dresden-II data~\cite{Colaresi:2022obx}.
Concerning new physics scenarios, we first explore unitarity violation in the lepton mixing matrix, a characteristic feature of low-scale seesaw mechanisms.
Moreover, we consider neutrino generalized interactions (NGI) with heavy mediators 
by taking into consideration  two different quenching factor models namely, the photoneutron (YBe) and iron-filter (Fef) QF reported by the Dresden-II Collaboration. \\[-.5cm]
 
The remainder of the paper is organized as follows: In Sec.~\ref{sec:Theory} we provide the necessary theoretical framework regarding \cevns processes.
We begin by discussing \cevns within the SM followed by a brief review of the unitarity violation and NGIs.
Section \ref{sec:results} contains the details of our Dresden-II simulated signal and the statistical analysis we have adopted as well as discussion of our results.
We finally summarize our concluding remarks in Sec. \ref{sec:conclusions}.

\section{Theoretical Framework}  
\label{sec:Theory}

In this section we discuss the \cevns differential cross section within the SM framework.
We also describe the necessary modifications for incorporating unitarity violation effects in active neutrino mixing,  NSIs or the most general NGIs.
For the latter cases, we only consider the effective new physics interactions resulting from integrating out the heavy mediators.

\subsection{\cevns within the SM} 
\label{subsec:SM_xsec}
The tree-level SM differential \cevns cross section with respect to the nuclear recoil energy $E_{nr}$ is expressed as\footnote{It should be emphasized that Eq.~\eqref{equn:SM_xsec} holds for adequately low momentum transfer such that the coherence criterion $\mathfrak{q}\leq1/R_A$~\cite{Papoulias:2018uzy} holds,
  where $R_A$ is the nuclear radius and $\mathfrak{q}^2=2m_NE_{nr}$ is the three-momentum transfer.}~\cite{Drukier:1984vhf}
\begin{equation}
\label{equn:SM_xsec}
\left.\frac{d\sigma}{dE_{nr}}\right|_{\text{SM}}=\frac{G_F^2m_N}{\pi}Q_W^2\left(1-\frac{m_NE_{nr}}{2E_\nu^2} - \frac{E_{nr}}{E_\nu}\right)\, ,
\end{equation} 
where $G_F$ represents the Fermi constant and $m_N$ the mass of the target nucleus. Here the SM vector weak charge $Q_W$ is explicitly written as~\cite{Barranco:2005yy},
\begin{equation}
\label{equn:SM_Charge}
Q_W= g_V^pZF_Z(\mathfrak{q}^2)+g_V^nNF_N(\mathfrak{q}^2) \,,
\end{equation}
where $Z$ and $N$ denote the number of protons and neutrons of the target nucleus, while $g_V^p~=1/2-2\sin^2{\theta_W}$ and $g_V^n=-1/2 $ denote the corresponding SM tree-level vector couplings for protons and neutrons, with $\theta_W$ representing the weak mixing angle. We have verified that the radiative corrections employed in \cite{AtzoriCorona:2022qrf} lead to less than 1\% difference in the expected event rates. 
Equation~\eqref{equn:SM_Charge} also incorporates the nuclear proton and neutron  form factors $F_Z(\mathfrak{q}^2)$ and $F_N(\mathfrak{q}^2)$,
which are the Fourier transforms of the corresponding nuclear charge density distribution and reflect the loss of coherence for sizable momentum transfer~\cite{Papoulias:2019xaw,Papoulias:2015vxa}.
The nuclear form factors are essential in the analysis of COHERENT data which utilized $\pi$-DAR neutrino beams.
Note, however, that for the recoil energy region of interest (ROI) at the Dresden-II experiment $F_Z(\mathfrak{q}^2)$ and $F_N(\mathfrak{q}^2)$ are roughly equal to unity.

\subsubsection*{Implications of weak mixing angle}
\label{sec:impl-weak-mixing}

The weak mixing angle is an important input of the Salam-Weinberg electroweak theory~\cite{Weinberg:1967tq}.
The theoretically estimated value of the weak mixing angle at the $Z$ pole is $\left.\sin^2\theta_W\right|_{\overline{\text{MS}}}(\mu=m_{Z^0})=0.23121\pm 0.00004$.  
For low energies, relevant to the \cevns process, i.e. $\mathfrak{q}\approx0$, its value has been obtained by RGE extrapolation in the minimal
subtraction ($\overline{\text{MS}}$) renormalization scheme: $\sin^2\theta_W(\mathfrak{q}=0)=0.23857\pm 0.00005$ \cite{10.1093/ptep/ptaa104, Erler:2019hds}.
A precise measurement of this parameter at low energies is an important test for the validation of the SM. 
Although many efforts have been conducted so far~\cite{10.1093/ptep/ptaa104}, measurements of the weak mixing angle at low energy still come up with significant uncertainties.  
\cevns is a low-energy process that can be used for the determination of the weak mixing angle in the low-energy regime. 
Indeed, Ref.~\cite{Cadeddu:2021ijh} pointed out that the uncertainty in the measurement of $\sin^2 \theta_W$ obtained from the combined fit of the full COHERENT-CsI + APV~\footnote{APV stands for atomic parity violation in cesium atoms; it is measured in Refs.~\cite{Wood:1997zq, Guena:2004sq}. }
datasets is still rather large compared to other determinations of this parameter at low momentum transfer.
Note that a limiting factor behind the poor COHERENT sensitivity on the weak mixing angle relates to the additional nuclear physics uncertainties involved in the \cevns signal~\cite{Cadeddu:2021ijh}.
On the other hand, being free from nuclear physics uncertainties, the Dresden-II reactor data offer a new opportunity for an improved determination of the weak mixing angle.

\subsection{Violation of lepton unitarity}  
\label{subsec:NU_Paramters}
Neutrinos in the SM are massless and have no isosinglet ``right-handed'' partners. 
A simple way to give them mass is the (high-scale) seesaw mechanism in which one postulates the existence of a right-handed neutrino associated with each left-handed one,
following in a conventional ``(3,3)'' structure.
Given that right-handed partners are gauge singlets one can add any number of them~\cite{Schechter:1980gr,Schechter:1981cv}. 
The low-scale seesaw mechanisms assume a common ``(3,6)'' seesaw template:
instead of having a single heavy isosinglet right-handed neutrino associated with each family,
it contains two heavy singlets in each family, making up three sequential Dirac leptons, in the limit of lepton-number conservation.  
  These schemes include both the inverse~\cite{Mohapatra:1986bd,Gonzalez-Garcia:1988okv} and the linear seesaw mechanism~\cite{Akhmedov:1995ip,Akhmedov:1995vm,Malinsky:2005bi}. 
 In both of them, a massless neutrino setup is the common template for building a genuine low-scale seesaw mechanism,
 in which the small symmetry-protected neutrino masses are mediated by heavy quasi-Dirac neutrino exchange. 
 These schemes can be probed directly at high energies~\cite{Dittmar:1989yg,Gonzalez-Garcia:1990sbd,AguilarSaavedra:2012fu,Das:2012ii,Deppisch:2013cya},
 as in the ATLAS and CMS collaborations at the LHC~\cite{ATLAS:2019kpx,CMS:2018iaf,CMS:2022nty},
 or in future neutral heavy lepton search proposals---see, e.g., Refs.~\cite{Drewes:2019fou,Abdullahi:2022jlv}.

 Low-scale seesaw schemes imply signals at low-energy precision studies of neutrino properties, such as unitarity violation in neutrino mixing. 
Indeed, a characteristic feature of these models is, of course, the existence of new neutral leptons.
The latter are assumed to be heavy and participate in weak interaction processes only through mixing with the SM (active) neutrinos.
The admixture of heavy leptons in the charged current implies that the weak interaction mixing matrix needs to be rectangular~\cite{Schechter:1980gr},
 leading to effective violation of lepton unitarity in many processes. \\[-.5cm] 

 Here we consider the possibility of constraining the nonunitarity (NU) parameters in the light of the Dresden-II reactor data. 
 The generalized charged-current weak interaction mixing matrix characterizing unitarity violation can be written in the form~\cite{Escrihuela:2015wra}
\begin{equation}
\label{equn:generalized_charged_current_weak_interaction_mixing_matrix}
N=N^{\text{NP}}U^{3\times 3}\, ,
\end{equation}
where $U^{3\times 3}$ is the standard $3 \times 3$ lepton mixing matrix, while $N^{\text{NP}}$ denotes the new physics matrix describing unitarity violation.
The matrix $N^{\text{NP}}$ can be conveniently parametrized as follows~\cite{Escrihuela:2015wra} 
\begin{equation}
\label{equn:New_Physics_Matrix}
N^{\text{NP}}\equiv \begin{pmatrix}
\alpha_{11}&0&0\\
\alpha_{12}&\alpha_{22}&0\\
\alpha_{13}&\alpha_{23}&\alpha_{33}
\end{pmatrix}\, ,
\end{equation}
where the diagonal entries ($\alpha_{ii}$) are real,  while the off-diagonal ones ($\alpha_{ij}$) are small but in general complex.

Within this context, the initial neutrino flux generated at the source, undergoes oscillation in propagation and reaches the detector as a modified flux according to~\cite{Miranda:2020syh}
\begin{equation}
\label{equn:Oscillation_Probability}
\begin{pmatrix}
\frac{d\Phi^\text{NU}_{\nu_e}}{dE_\nu}+\frac{d\Phi^\text{NU}_{\bar{\nu}_e}}{dE_\nu}\\
\frac{d\Phi^\text{NU}_{\nu_\mu}}{dE_\nu}+\frac{d\Phi^\text{NU}_{\bar{\nu}_\mu}}{dE_\nu}\\
\frac{d\Phi^\text{NU}_{\nu_\tau}}{dE_\nu}+\frac{d\Phi^\text{NU}_{\bar{\nu}_\tau}}{dE_\nu}\\
\end{pmatrix}=\begin{pmatrix}
P_{ee}&P_{e\mu}&P_{e\tau}\\
P_{e\mu}&P_{\mu\mu}&P_{\mu\tau}\\
P_{e\tau}&P_{\mu\tau}&P_{\tau\tau}
\end{pmatrix}\begin{pmatrix}
\frac{d\Phi_{\nu_e}}{dE_\nu}+\frac{d\Phi_{\bar{\nu}_e}}{dE_\nu}\\
\frac{d\Phi_{\nu_\mu}}{dE_\nu}+\frac{d\Phi_{\bar{\nu}_\mu}}{dE_\nu}\\
\frac{d\Phi_{\nu_\tau}}{dE_\nu}+\frac{d\Phi_{\bar{\nu}_\tau}}{dE_\nu}\\
\end{pmatrix}\, ,
\end{equation}
where $P_{\alpha\beta}$ represents the probability of transition from $\alpha$ to $\beta$ neutrino or antineutrino flavor ($P_{\alpha\beta}=P[\nu_\alpha\to\nu_\beta]=P[\bar{\nu}_\alpha\to\bar{\nu}_\beta]$) and satisfies the condition $P_{\alpha\beta}=P_{\beta\alpha}$.
The oscillation probabilities in the presence of NU are expressed as~\cite{Escrihuela:2015wra}
\begin{equation}
\begin{aligned}
&P_{ee}=\alpha_{11}^4P_{ee}^{3\times3}\, ,\\
\ignore{&P_{\mu\mu}=\alpha_{22}^4P_{\mu\mu}^{3\times3}+\alpha_{22}^3\left|\alpha_{12}\right|P_{\mu\mu}^{I_{1}}+2\left|\alpha_{12}\right|^2\alpha_{22}^2P_{\mu\mu}^{I_2}\, ,\\}
&P_{e\mu}=\left(\alpha_{11}\alpha_{22}\right)^2P_{e\mu}^{3\times 3}+\alpha_{11}^2\alpha_{22}\left|\alpha_{21}\right|P_{e\mu}^I+\alpha_{11}^2\left|\alpha_{12}\right|^2\, , \\
&P_{e\tau}=\left(\alpha_{11}\alpha_{33}\right)^2P_{e\tau}^{3\times 3}+\alpha_{11}^2\alpha_{33}\left|\alpha_{31}\right|P_{e\tau}^I+\alpha_{11}^2\left|\alpha_{13}\right|^2\, .
\end{aligned}
\end{equation}
where $P^{3\times3}_{ee}$ is the standard survival probability, and $P^{3\times3}_{e\mu}$ and $P^{3\times3}_{e\tau}$ are the standard transition probabilities, while $P_{e\mu}^I$ and $P_{e\tau}^I$ contain CP-violating terms as explained in Ref.~\cite{Escrihuela:2015wra}.  
For very short baseline experiments such as Dresden-II, the standard neutrino oscillation probability can be well approximated as $P_{\alpha\beta}^{3\times 3}(L \approx 0)=\delta_{\alpha\beta}$, with $\delta_{\alpha\beta}$ denoting the Kronecker delta (zero-distance effect). 
It follows that the expected number of events can be cast in the simple form 
\begin{equation}
R_{\text{NU}} = \alpha_{11}^2 \left(\alpha_{11}^2 + |\alpha_{12}|^2 +   |\alpha_{13}|^2 \right) R_{\text{SM}} \, ,
\label{eq:events_NU}
\end{equation}
where $R_{\text{SM}}$ represents the unoscillated number of \cevns events in the SM.
Hence, for the case of reactor $\bar{\nu}_e$, the only relevant NU parameters affecting the propagation of neutrinos are $a_{11}$, $|\alpha_{12}|$, and $|\alpha_{13}|$.
Note that, in contrast to oscillation experiments, the expected neutrino signal in short-baseline reactor neutrino experiments is directly proportional to the NU parameters, independently of CP-violating terms and standard oscillation phenomena.
To our knowledge,  up to now, only projected NU sensitivities for the \cevns channel exist in the literature---e.g., using future data expected at COHERENT~\cite{Miranda:2020syh} or at the planned SBC experiment~\cite{Alfonso-Pita:2022eli}.
In the present work, we instead consider the actual data from the Dresden-II experiment. 

\subsection{Neutrino nonstandard interactions}
\label{subsec:NSI_xsec}

Neutrino nonstandard interactions (or NSIs, for short)~\cite{Wolfenstein:1977ue,Valle:1987gv} arise in most neutrino mass generation schemes, and
their expected magnitude can be sizeable in the case of low-scale models~\cite{Boucenna:2014zba}.
Prominent theories of neutrino mass generation borrow the chiral structure of the \sm and, as a result, imply a restricted Lorentz structure for the resulting NSIs.
These have been extensively studied in various contexts using Wilson coefficients of nonrenormalizable dimension-6 operators~\cite{Lee:1956qn, Ohlsson:2012kf, Farzan:2017xzy, Billard:2018jnl}.
NSIs have been previously searched for using the COHERENT data~\cite{COHERENT:2017ipa, COHERENT:2021xmm,COHERENT:2020iec}
(see Refs.~\cite{Papoulias:2017qdn, Papoulias:2019txv, Miranda:2020tif, Giunti:2019xpr,Khan:2021wzy,Dutta:2020che, Canas:2019fjw, Coloma:2019mbs})
and the CONUS data~\cite{CONUS:2021dwh}.
Studies have also been performed simulating future data expected at the European Spallation Source~\cite{Chatterjee:2022mmu} and at the SBC~\cite{Alfonso-Pita:2022eli}. 

Because of the basic gauge structure NSIs are restricted to be vector or axial vector-type interactions (see below the most general case of the NGIs).
The relevant nonstandard-interaction neutral-current Lagrangian reads~\cite{Barranco:2005yy} 
\begin{equation}
\mathscr{L}^{\text{NSI}}_{\text{eff}} \;=\;
-2\sqrt{2}\,G_F\, 
\varepsilon_{\alpha\beta}^{qC}\,\bigl(\overline{\nu}_\alpha\gamma^\mu P_L \nu_\beta\bigr)
\bigl(\overline{q}\gamma_\mu P_C q\bigr)
%+ h.c.
\;,
\label{H_NC-NSI}
\end{equation}
where $\alpha, \beta = e,\mu,\tau$ is the neutrino flavor,  $q = u,d$ stands for the quark fields, and $C=L, R$ denotes the chirality projections. 
The strength of the new interaction is taken with respect to the Fermi constant, through the NSI parameters $\varepsilon_{\alpha\beta}^{qC}$,
which can be either flavor preserving ($\alpha =\beta$) or flavor changing ($\alpha \neq \beta$).
The former, flavor-preserving NSIs, are usually referred to as nonuniversal. 
It is often convenient to express the latter in vector and axial-vector form as
$\varepsilon_{\alpha\beta}^{qV} \equiv \varepsilon_{\alpha\beta}^{qR} + \varepsilon_{\alpha\beta}^{qL}$ and
$\varepsilon_{\alpha\beta}^{qA} \equiv \varepsilon_{\alpha\beta}^{qR} - \varepsilon_{\alpha\beta}^{qL}$, respectively. \\[-.5cm]

For the case of \cevns on intermediate or heavy target nuclei, axial-vector NSIs are usually ignored, since they are expected to contribute
negligibly\footnote{Axial-vector \cevns interactions are suppressed by a factor $\sim 1/A$ (see Ref.~\cite{Barranco:2005yy}).}
with respect to their vector counterpart. 
As a result, for an incoming (anti)neutrino of flavor $\alpha$, the NSI \cevns cross section is given from Eq.(\ref{equn:SM_xsec})
with the substitution $Q_{W} \to Q_{\text{NSI}}^V$, where the NSI charge reads
\begin{eqnarray}
Q_{\text{NSI}}^V & = \left(g_V^p + 2\varepsilon_{\alpha\alpha}^{uV}+\varepsilon_{\alpha\alpha}^{dV}\right) Z F_Z(\mathfrak{q}^2) + \left(g_V^n + \varepsilon_{\alpha\alpha}^{uV}+2\varepsilon_{\alpha\alpha}^{dV}\right) N F_N(\mathfrak{q}^2)  \nonumber \\
 & + \sum\limits_{\beta\neq \alpha}  \left(2 \varepsilon_{\alpha\beta}^{uV} + \varepsilon_{\alpha\beta}^{dV} \right)  Z F_Z(\mathfrak{q}^2) + \left(\varepsilon_{\alpha\beta}^{uV} + 2\varepsilon_{\alpha\beta}^{dV} \right) N F_N(\mathfrak{q^2})  \, ,
 \label{eq:weak_charge_NSIs}
\end{eqnarray}
with $g_V^p$ and $g_V^n$ denoting the corresponding SM vector couplings for protons and neutrons [see Eq.(\ref{equn:SM_Charge})].

\subsection{Neutrino generalized interactions}
\label{subsec:NGI_xsec}

Generalized neutrino interactions provide a model-independent framework for a variety of new physics scenarios. 
They incorporate all possible Lorentz-invariant interactions up to dimension $6$ between neutrinos and other SM fermions---i.e., scalar, pseudoscalar, vector, axial vector, and tensor~\cite{Lindner:2016wff}---hence, extending the conventional NSIs discussed above. 

Effective field theories (EFTs) resulting from new physics~\cite{Jenkins:2017jig} provide a suitable framework for describing NGIs in a model-independent manner below the electroweak scale. 
They can be described through the four-fermion effective interaction Lagrangian~\cite{AristizabalSierra:2018eqm}
\begin{equation}
\label{equn:NGI+quark_Lagrangian}
\mathscr{L}^{\text{NGI}}_{\text{eff}}=\frac{G_F}{\sqrt{2}}\sum_{X=S,P, V,A, T}\left[\bar{\nu}\Gamma^X \nu\right]\left[\bar{q}\Gamma_X\left(C_X^{q}+i\gamma_5D_X^{q}\right) q\right]\, ,
\end{equation}
with $\Gamma_X=\left\{\mathbb{1},\gamma_5, \gamma_\mu,\gamma_\mu\gamma_5, \sigma_{\mu\nu}\right\}$ corresponding to $X=\{S,P,V,A,T\}$ interactions.

The relative strength of the new physics interaction $X$ is specified by the real dimensionless coefficients $C_X^q$ and $D_X^{q}$ which are of the order $(\sqrt{2}/G_F)(g_X^{\nu}g_X^{q}/m_X^2)$.
Here, $m_X$ denotes the mediator mass $(\mathfrak{q} \ll m_X)$, and $g_X^\nu$ ($g_X^q$) denotes the coupling between the mediator and neutrino (quark). 
Note that we have neglected pseudoscalar and axial vector interactions, since they are expected to be suppressed by the nuclear spin; we choose $D_X^{q}=0$.

The transition from quark-level to nuclear-level interactions is achieved by assuming that the nucleonic matrix element of the quark current at $\mathfrak{q}=0$ is proportional to that of the corresponding nucleon current---i.e., $\left<N_f|\bar{q}\Gamma^X q|N_i\right>=\mathcal{F}^X\left<N_f|\bar{N}\Gamma^X N|N_i\right>$ ($\mathcal{F}^X$ is the hadronic form factor) as argued in Refs.~\cite{AristizabalSierra:2018eqm, Dent:2015zpa, Cirelli:2013ufw}. The neutrino-nucleus effective Lagrangian reads~\cite{AristizabalSierra:2018eqm} 
\begin{equation}
\label{equn:NGI+nuclear_Lagrangian}
\mathscr{L}_{\nu -N}\sim \sum_{X=S,V,T} C_X\left[\bar{\nu}\Gamma^X\nu\right]\left[\bar{N}\Gamma_XN\right]\, ,
\end{equation}
where the neutrino-nucleus effective couplings $C_X$ are expressed in terms of $C_X^q$ as~\cite{AristizabalSierra:2018eqm} 
\begin{gather}
\begin{aligned}
C_S=& Z \, F_Z(\mathfrak{q}) \, \sum_q C_S^q\frac{m_p}{m_q}f^p_{q}+N \, F_N(\mathfrak{q})\sum_q C_S^q\frac{m_n}{m_q}f^n_{q} \, ,\\
C_{V}=&  Z \, F_Z(\mathfrak{q}) \left(2C_V^u+C_V^d\right)+ N \, F_N(\mathfrak{q})\left(C_V^u+ 2C_V^d \right)\, ,\\
C_T=&  Z \, F_Z(\mathfrak{q})\, \sum_q C_T^q\delta^p_{q}+N \, F_N(\mathfrak{q})\, \sum_q C^q_T\delta^n_{q} \, .
\end{aligned}
\label{eq:BSM_charges_cevns}
\end{gather}
The hadronic form factors for the scalar case ($S$) are $f_u^p=0.0208$, $f_u^n=0.0189$, $f_d^p=0.0411$,  and $f_d^n=0.0451$~\cite{AristizabalSierra:2019zmy},
and those for the tensor case ($T$) are $\delta^p_u=\delta^n_d=0.54$  {and} $\delta^p_d=\delta^n_u=-0.23$ \cite{AristizabalSierra:2019zmy}.  

Neglecting the higher-order terms of $\mathcal{O}(E_{nr}^2/E_\nu^2)$ and taking into account the simultaneous presence of all these interactions
the differential cross section corresponding to the Lagrangian density in Eq.\eqref{equn:NGI+nuclear_Lagrangian} takes the form~\cite{AristizabalSierra:2018eqm} 
\begin{equation}
\begin{aligned}
\left.\frac{d\sigma}{dE_{nr}}\right|_{\text{NGI}}=\frac{G_F^2m_N}{\pi}\Bigl\{&C_S^2 \frac{m_NE_{nr}}{8E_\nu^2}+\left(\frac{C_V}{2}+Q_W\right)^2\left(1-\frac{m_NE_{nr}}{2E_\nu^2}-\frac{E_{nr}}{E_\nu}\right)\\
&+2C_T^2\left(1-\frac{m_NE_{nr}}{4E_\nu^2}-\frac{E_{nr}}{E_\nu}\right)\pm \mathcal{R}\frac{E_{nr}}{E_\nu} \Bigr\}\, . 
\end{aligned}
\label{equn:NGI_xSec}
\end{equation}

In the above expression, one notices the interference between the SM and vector NGIs, as well as the interference between the scalar and tensor terms given as $\mathcal{R} = C_SC_T/2$.  
The plus (minus) sign accounts for coherent elastic antineutrino (neutrino) scattering off nuclei (see also Ref.~\cite{Flores:2021kzl}). 
%%%
Notice that only flavor-diagonal NGI couplings have been considered in Eq.(\ref{equn:NGI_xSec}), while nondiagonal NGI coefficients are neglected. 
%%%
It is worth mentioning also that the vector NSI contributions discussed in Sec.~\ref{subsec:NSI_xsec} are accounted for in the vector part of the generic NGI cross section,
since $C_V^{u} \equiv 2\varepsilon^{uV}_{\alpha \alpha}$ and $C_V^{d} \equiv 2\varepsilon^{dV}_{\alpha \alpha}$.  

NGIs have been previously searched for through \cevns using the COHERENT CsI measurement data~\cite{AristizabalSierra:2018eqm},
while projected sensitivities at CONUS~\cite{Lindner:2016wff} and the planned SBC experiment have also been reported~\cite{Flores:2021kzl}. 
Here, for the first time we explore the NGIs parameter space using actual reactor data, from the Dresden-II experiment.  
Notice also the complementarity of our present work with Ref.~\cite{AristizabalSierra:2022axl}, which analyzed the Dresden-II data assuming light mediators.

\section{Results}
\label{sec:results}

In this section we briefly review the specifications of the Dresden-II experiment~\cite{Colaresi:2021kus, Colaresi:2022obx}, and provide the technical details required to successfully simulate the \cevns and background signals expected at the NCC-1701 detector. We also describe the procedure of the statistical analysis adopted in this work and discuss our results. 

\subsection{Simulation of the Dresden-II signal}
\label{subsec:events_spectra}

In its current configuration, the Dresden-II reactor experiment used a $2.924$ kg ultralow-noise p-type point-contact germanium detector.
It was exposed to the intense electron antineutrino flux coming from the Dresden-II boiling water reactor.
The experiment has collected data for 96.4 days (Rx-ON) during which the reactor operated with its full nominal power of $P=2.96$~GW.  
The antineutrino energy flux used here is taken from Appendix~A of Ref.~\cite{CONNIE:2019xid},
\begin{equation}
\frac{d \Phi_{\bar{\nu}_e}}{d E_\nu} = \frac{P}{4 \pi d^2 \epsilon} \, \sum_{i} \frac{d N^i_{\bar{\nu}_e}}{d E_\nu}\, ,
\end{equation}
where $\epsilon=205.24$~MeV is the average energy released per fission and the index $i$ runs over all the fissible
isotopes $\mathrm{^{235}U}$, $\mathrm{^{238}U}$, $\mathrm{^{239}Pu}$, $\mathrm{^{241}Pu}$ and $\mathrm{^{238}U(n, \gamma)^{239}U}$.
Given the baseline of the experiment $d=10.39$~m, the neutrino flux normalization reaching the NCC-1702 detector can be estimated to be $\mathscr{N}=  4.8\times 10^{13}/\text{cm}^2/\text{s}$.

The theoretical CE$\nu$NS events in the $i$th bin can be estimated as\footnote{The nuclear recoil energy spectra has a lower cutoff ($E_{nr}^\text{min}$) corresponding to electron equivalent ionization energy, $E_{er,\text{ min}}^\text{true}\simeq 2.98\text{ eV}_{ee}$, necessary to produce a electron-hole pair in Ge at $77$~K~\cite{ANTMAN1966272, Wei:2016xbw}.}~\cite{AristizabalSierra:2022axl}
\begin{equation}
\label{equn:CEvNS_Events}
\left[R_i\right]_\xi^{\text{CE}\nu\text{NS}}=t_{\text{run}}N_{\text{target}}\int_{E_{er}^i}^{E_{er}^{i+1}}dE_{er}^{\text{reco}}\int_{E_{nr}^\text{min}}^{E_{nr}^\text{max}} dE_{nr} \, G(E_{er}^{\text{reco}}, E_{er}^{\text{true}})\int_{E_\nu^\text{min}}^{E_\nu^\text{max}}dE_\nu \frac{d\Phi_{\bar{\nu}_e}}{dE_\nu}\left.\frac{d\sigma}{dE_{nr}}\right|_\xi\, ,
\end{equation}
where $\xi = \{\text{SM, NU, NGI} \}$ denotes the type of \cevns interaction\footnote{Note that, by switching off the scalar ($C_S$) and tensor ($C_T$) terms in Eq.(\ref{equn:NGI_xSec}), the NGI cross section reduces to that for the conventional NSI \cevns case. From here on, we give the NSI results as a particular case of the NGI ones.}.
Here, $t_{\text{run}}$ represents the exposure time, and $N_{\text{target}}=m_\text{det}N_A/m_\text{Ge}$ denotes the number of target nuclei.
$N_A$ is the Avogadro number, $m_\text{Ge}$ the molar mass of the Ge isotope, and $m_\text{det}=2.924$~kg the mass of NCC-1701. We consider all the stable isotopes of Ge with mass number $\{70,72,73,74,76\}$ along with their corresponding relative abundances $\{0.2057, 0.2745, 0.0775, 0.3650, 0.0773\}$ taken from~\cite{BerglundWieser:2011}.
In Eq. \eqref{equn:CEvNS_Events}, $E_\nu^\text{max}$ ($E_\nu^\text{min}$) denotes the maximum (minimum) electron antineutrino energy,
while the kinematics of the process is used to determine the $E_\nu^\text{min}$ required to produce a nuclear recoil with energy $E_{nr}$, as 
\begin{equation}
\label{equn:Min_v_Energy}
E_\nu^\text{min}=\frac{1}{2}\left[E_{nr}+\sqrt{E_{nr}^2+2m_NE_{nr}}\right]\approx\sqrt{\frac{m_NE_{nr}}{2}}\, .
\end{equation}
The ``electron equivalent'' ionization energy, $E_{er}^{\text{true}}$, is related to the nuclear recoil energy, $E_{nr}$, through the QF which quantifies the fraction of energy loss to heat,
as $E_{er}^{\text{true}}(E_{nr})=\mathcal{Q}_f(E_{nr})\cdot E_{nr}$. 
For the latter, we take into consideration the two QF models described in the Dresden-II data release~\cite{Colaresi:2022obx}.  
The first one (YBe) was obtained from observations of photoneutron sources~\cite{Scholz:2016qos, Collar:2021fcl},
while the second one (Fef) was determined from iron-filtered monochromatic neutrons.
A truncated Gaussian distribution is used to define the detector's energy-resolution function as~\cite{Coloma:2022avw}  
\begin{equation}
\label{equn:Resolution_Func}
G(E_{er}^{\text{reco}}, E_{er}^{\text{true}})=\left(\frac{2}{1+\text{Erf}\left(\frac{E_{er}^{\text{true}}}{\sqrt{2}\sigma_e}\right)}\right)  \frac{1}{\sqrt{2\pi}\sigma_e}  e^{-{\left(\frac{{E_{er}^{\text{reco}}-E_{er}^{\text{true}}}}{\sqrt{2}\sigma_e}\right)}^2}\, ,
\end{equation}
with $E_{er}^{\text{reco}}$ being the measured (or ``reconstructed") ionization energy. The energy resolution power of the detector has been determined to be $\sigma_e(E_{er}^{\text{true}})=\sqrt{\sigma_n^2+\eta \, F_f \, E_{er}^{\text{true}}}$.
Here, $\sigma_n=68.5~\mathrm{eV_{ee}}$ is the intrinsic electronic noise of the detector  (from Rx-ON data), and $\eta=2.96~\mathrm{eV_{ee}}$ denotes the average energy for electron-hole formation
in Ge, while $F_f=0.105$ is the average value of the germanium Fano factor~\cite{Colaresi:2022obx}.  

In the ROI of Dresden-II: $0.2~\text{keV}_{ee}\leq E_{er}^\text{reco}\leq 1.5\text{ keV}_{ee}$, the background model given in Ref.~\cite{Colaresi:2022obx} depends on four individual components,
coming from epithermal neutrons and three electron capture (EC) peaks and can be fully characterized by seven free parameters (see below).
Explicitly, the background model can be described as a constant term, a decaying exponential, and three Gaussian probability density functions (each depending on three free parameters:
amplitude, centroid, and standard deviation)~\cite{Colaresi:2022obx}:  
\begin{equation}
\label{equn:bkg}
R_\text{bkg}(\mathbf{\upbeta})=N_\text{epith}+A_\text{epith}\, e^{-\left(\frac{E_{er}^\text{reco}-E_\text{epith}}{\tau_\text{epith}}\right)}+\sum_{i=L_1,L_2,M}A_i \, e^{-{\left(\frac{E_{er}^\text{reco}-E_i}{\sqrt{2}\sigma_i}\right)}^2}\, .
\end{equation}
The elastic scattering of epithermal neutrons contributes at the same recoil energy region where the \cevns signal amounts to a significant number of events. 
This background component comprises a free constant term $N_\text{epith}$ plus a free decaying exponential function 
with amplitude $A_\text{epith}$, centroid $E_\text{epith} =0.2~\mathrm{keV_{ee}}$, and a decay constant $\tau_\text{epith}$.  
The remaining background contribution is due to the EC peaks in $^{71}\text{Ge}$, namely the $L_1$-, $L_2$-, and $M$-shell peaks. 
The $L_1$-shell peak depends on its own amplitude, centroid, and standard deviation $(A_{L_1}$, $E_{L_1}$, $\sigma_{L_1}$),
while the $L_2$- and $M$-shell peaks are expressed in terms of the $L_1$-shell peak parameters as given in Table~\ref{table:bkg_parameters}. 
The ratio of the number of counts under the $M$- and $L_1$-shell peaks, $\beta_{M/L_1}$, has been experimentally determined to be $0.16\pm0.03$~\cite{SchoenfeldSep1998, SuperCDMS:2015eex};
thus, the amplitude of the $M$-shell can be expressed as $A_M = \beta_{M/L_1}\times\frac{\sigma_{L_1}}{\sigma_M}\times A_{L_1}$, while its centroid is fixed to the nominal value
  $E_M = 0.158~\mathrm{keV_{ee}}$ as it can not be determined by the data. Finally its uncertainty is equal to the electronic noise $\sigma_M \equiv \sigma_n = 68.5~\mathrm{eV_{ee}}$~\cite{Colaresi:2022obx}. For the $L_2$ peak, the theoretical amplitude $A_{L_2}= 0.008 \times A_{L_1}$ and the centroid $E_{L_2} = 1.142~ \mathrm{keV_{ee}}$ are used, while its width is taken to be equal to that of the $L_1$-shell, i.e., $\sigma_{L_2} = \sigma_{L_1}$~\cite{Colaresi:2022obx}.
In what follows, each analysis requires fitting of the seven free background parameters, which are represented by the vector
$\mathbf{\upbeta}=\{N_\text{epith}, A_\text{epith}, \tau_\text{epith}, A_{L_1}, E_{L_1}, \sigma_{L_1}, \beta_{M/L_1}\}$. 
More details on the background treatment are given in the Appendix~\ref{sec:appendix}.
\begin{table}[t]
\centering
\begin{tabular}{|c|c|c|}
\hline
Parameters & \hspace{3mm}$L_2$-shell\hspace{3mm}& \hspace{3mm}$M$-shell\hspace{3mm} \\
\hline
Amplitude ($A_{i}$) & $0.008\times A_{L_1}$ & $\beta_{M/L_1}\times\frac{\sigma_{L_1}}{\sigma_M}\times A_{L_1}$\\
Centroid ($E_{i}$) & $1.142~\mathrm{keV_{ee}}$ & $0.158~\mathrm{keV_{ee}}$\\
Standard deviation ($\sigma_{i}$) & $\sigma_{L_1}$ & $68.5~\mathrm{eV_{ee}}$\\
\hline
\end{tabular}
\caption{Parameter details of $L_2$- and $M$-shell peaks. $\beta_{M/L_1}$ is the ratio of the number of counts under $M$- and $L_1$-shell peaks (see the main text and Ref.~\cite{Colaresi:2022obx} for details).}
\label{table:bkg_parameters}
\end{table}

\begin{figure}[t]
\begin{center}
\includegraphics[width=0.49\textwidth,height=5.5cm]{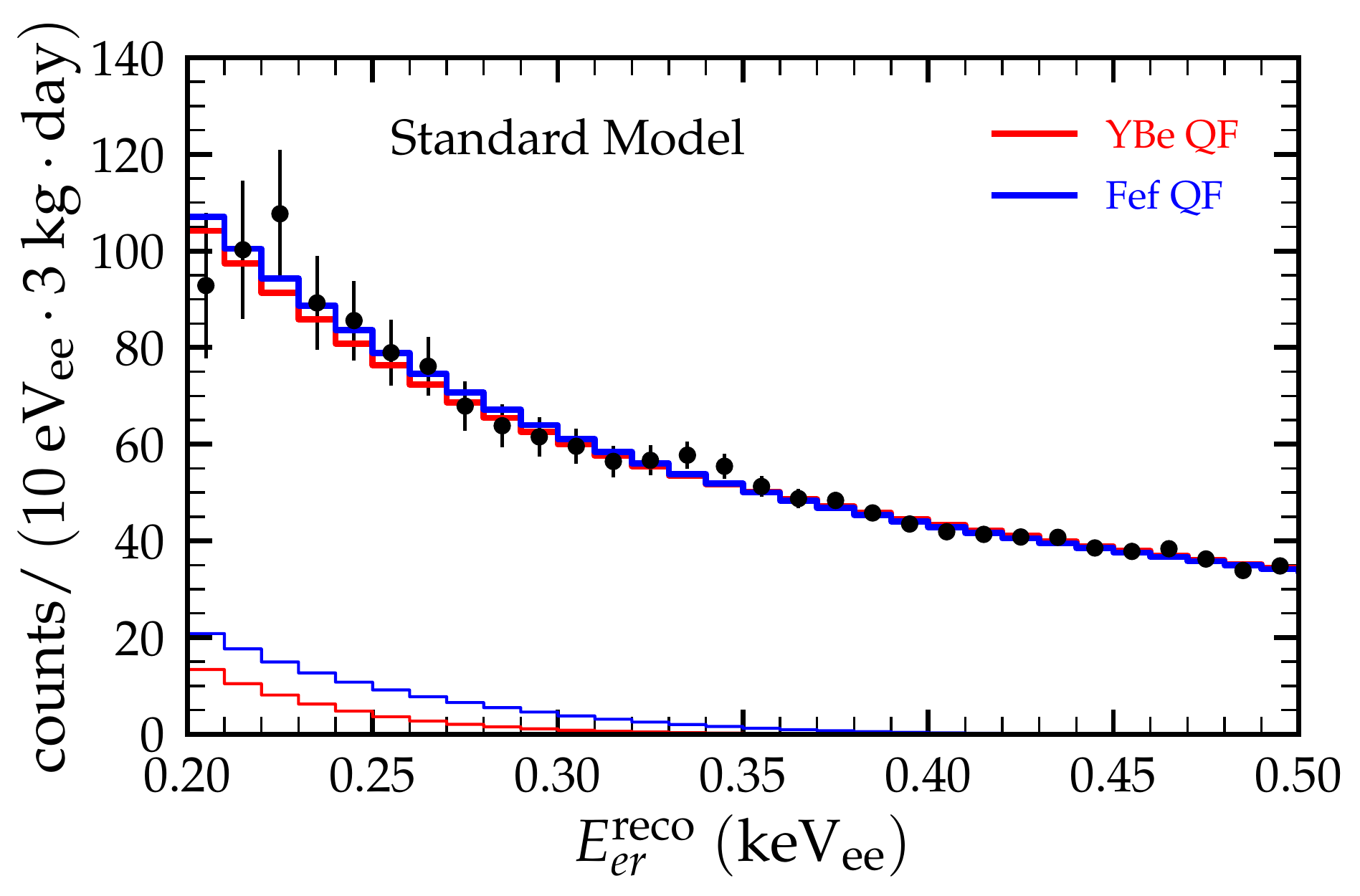}
\includegraphics[width=0.49\textwidth,height=5.5cm]{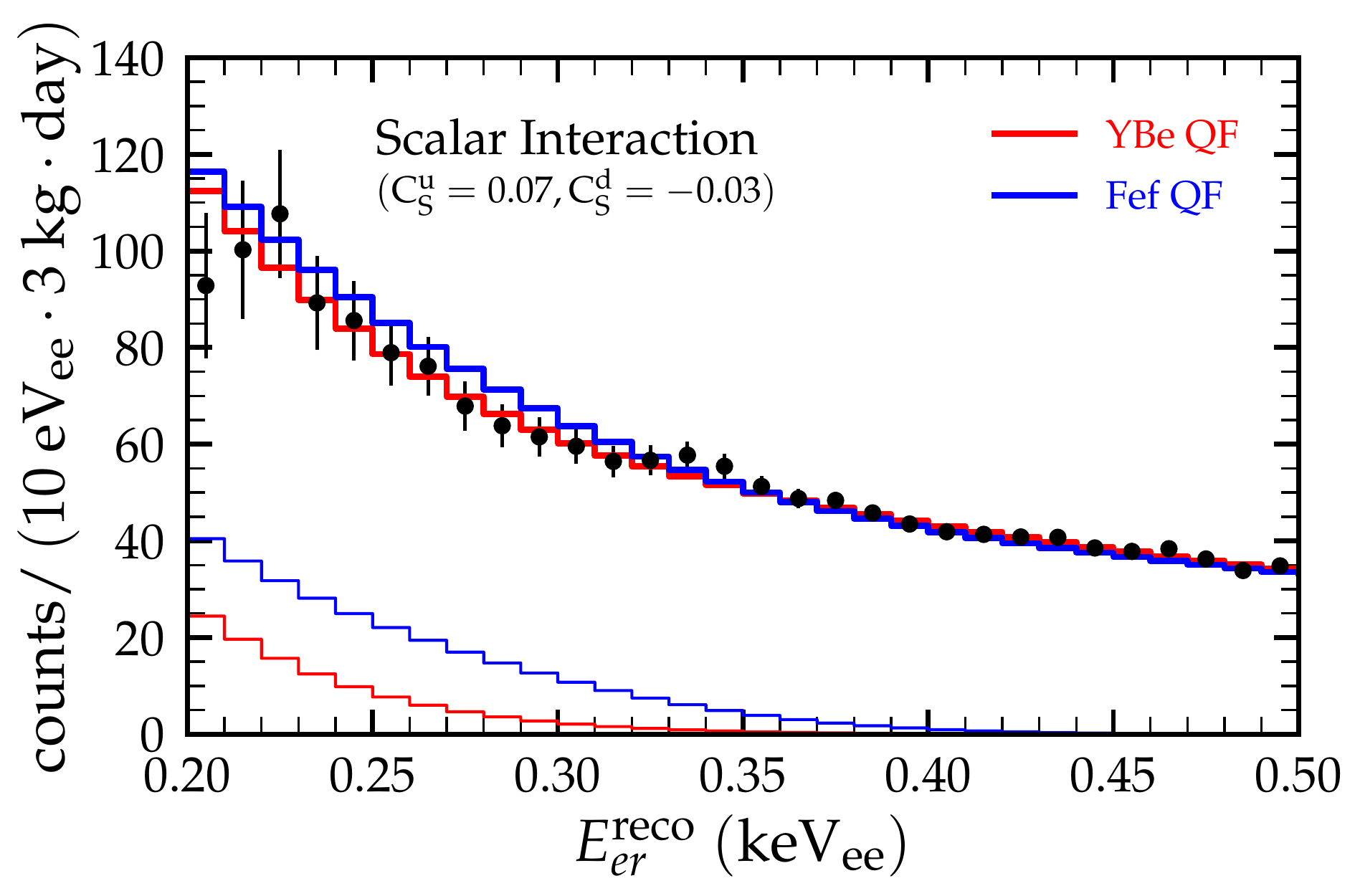}
\includegraphics[width=0.49\textwidth,height=5.5cm]{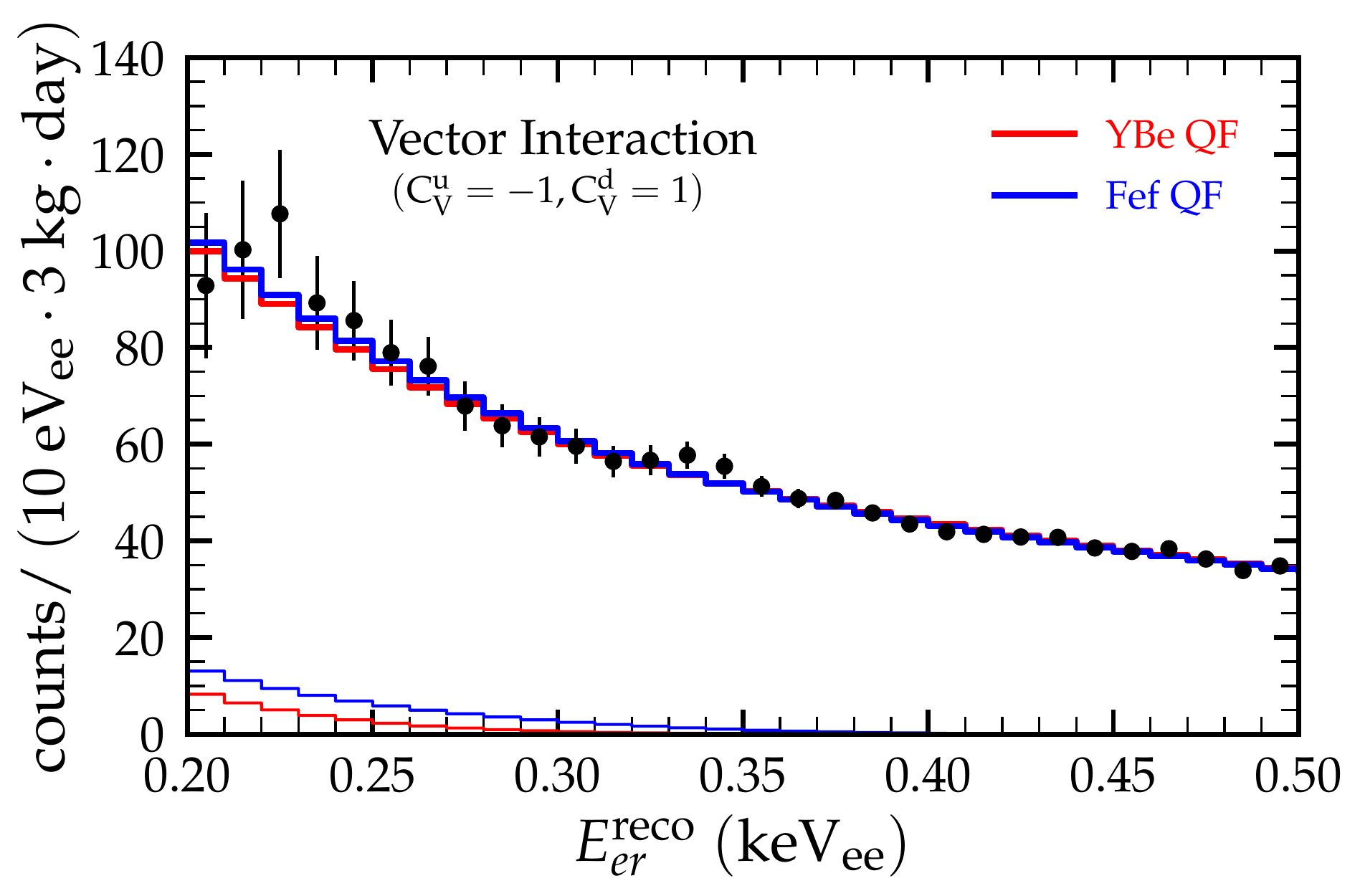}
\includegraphics[width=0.49\textwidth,height=5.5cm]{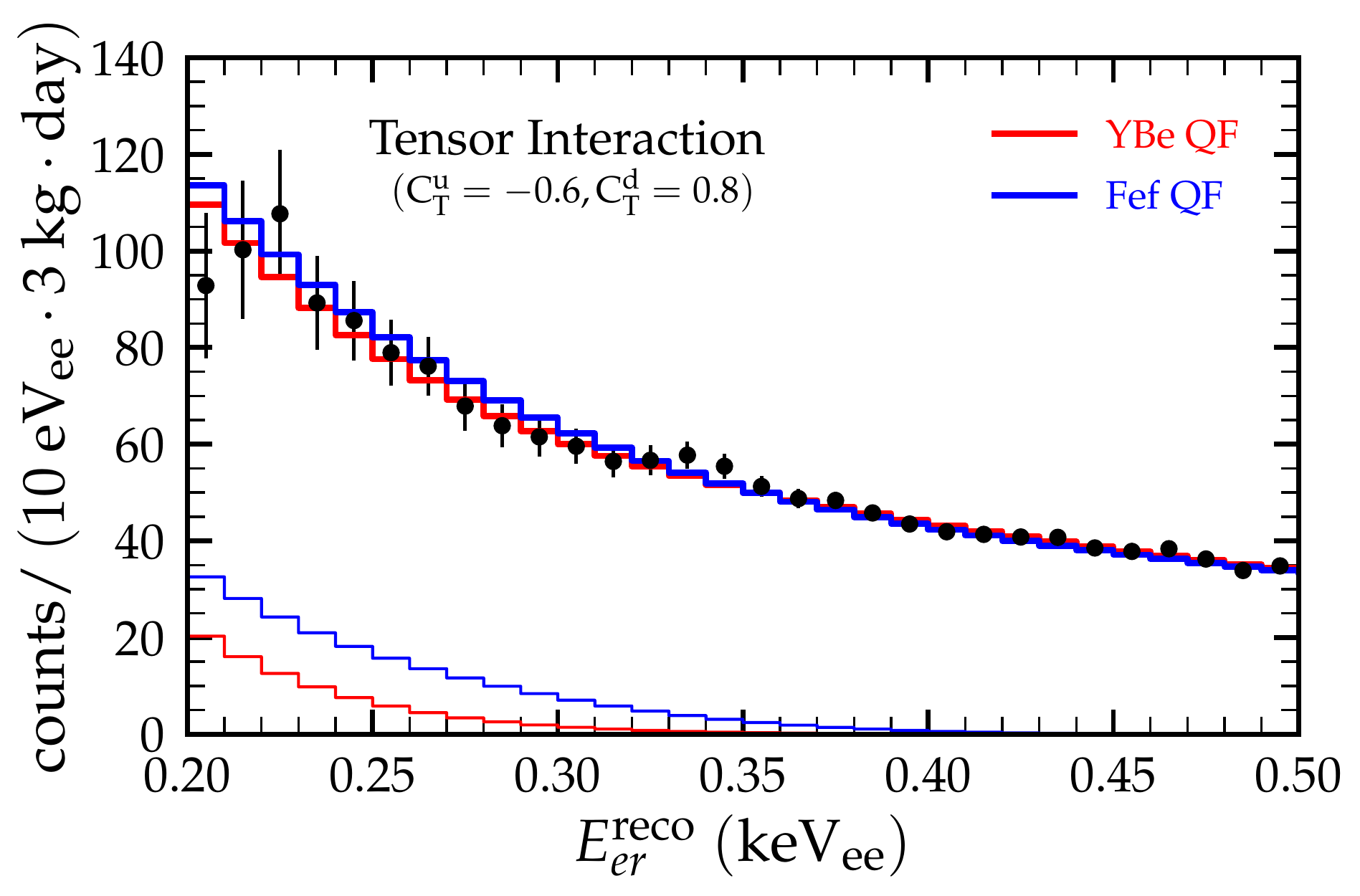}
\end{center}
\caption{\label{fig:Events} Estimated number of events within the SM and various new physics scenarios considered here, for given benchmark values for the new physics parameters.
  We also show a comparison of the Rx-ON data reported by the Dresden-II Collaboration. In each panel, the low-lying spectra with thin lines show the predicted signal from ``\cevns only,''
  while those with thick lines give the estimated ``\cevns + fitted background.'' Different colors correspond to two QF models.}
\end{figure}
\begin{figure}[t]
\begin{center}
\includegraphics[width=0.49\textwidth]{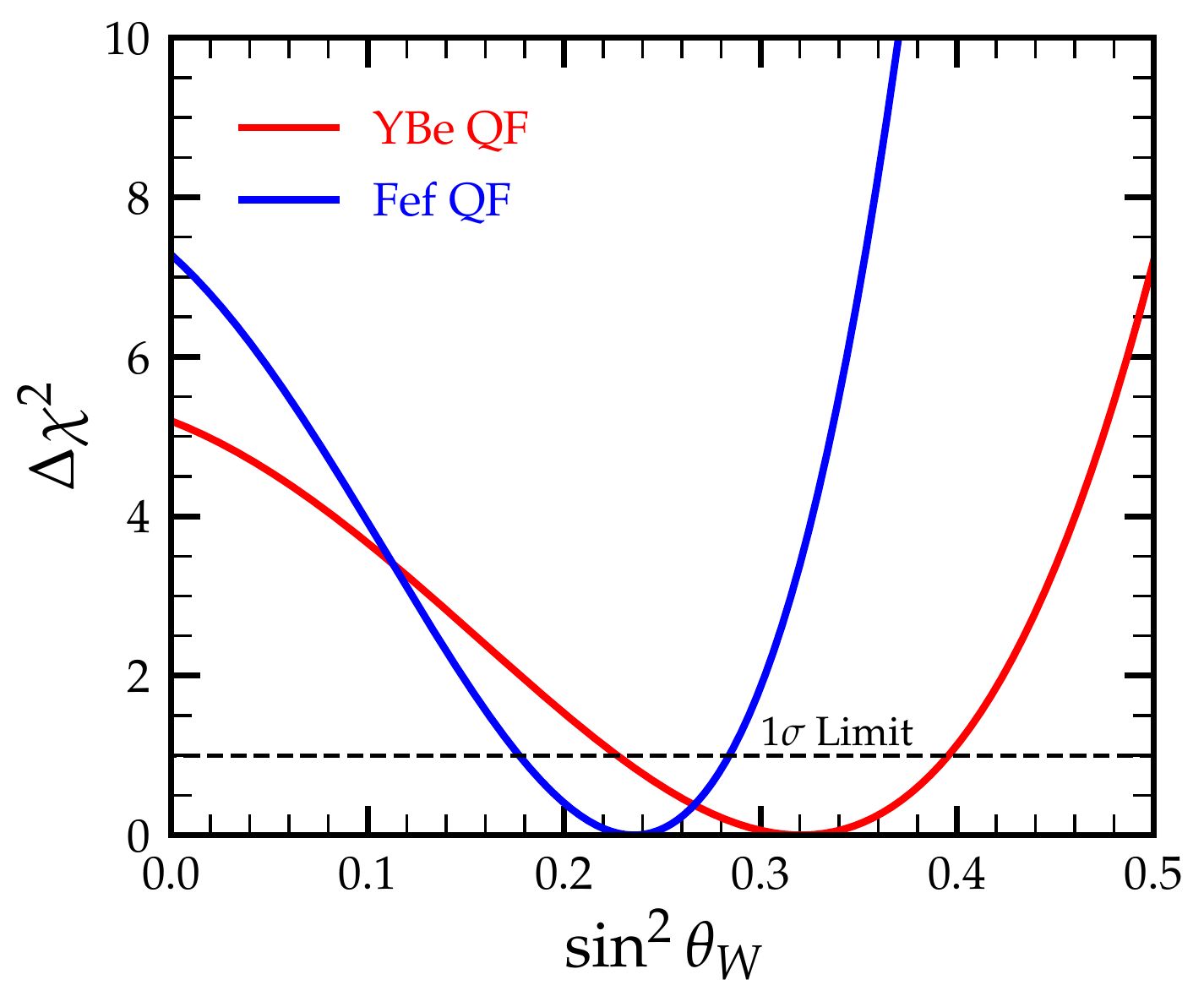}
\end{center}
\caption{Dresden-II sensitivity on the weak mixing angle, assuming the Fef and YBe QF models.}
\label{fig:Wienberg_Angle}
\end{figure}
%
%--------------------------------------------------------------------
\begin{figure}[t]
\begin{center}
  \includegraphics[width=0.8\textwidth]{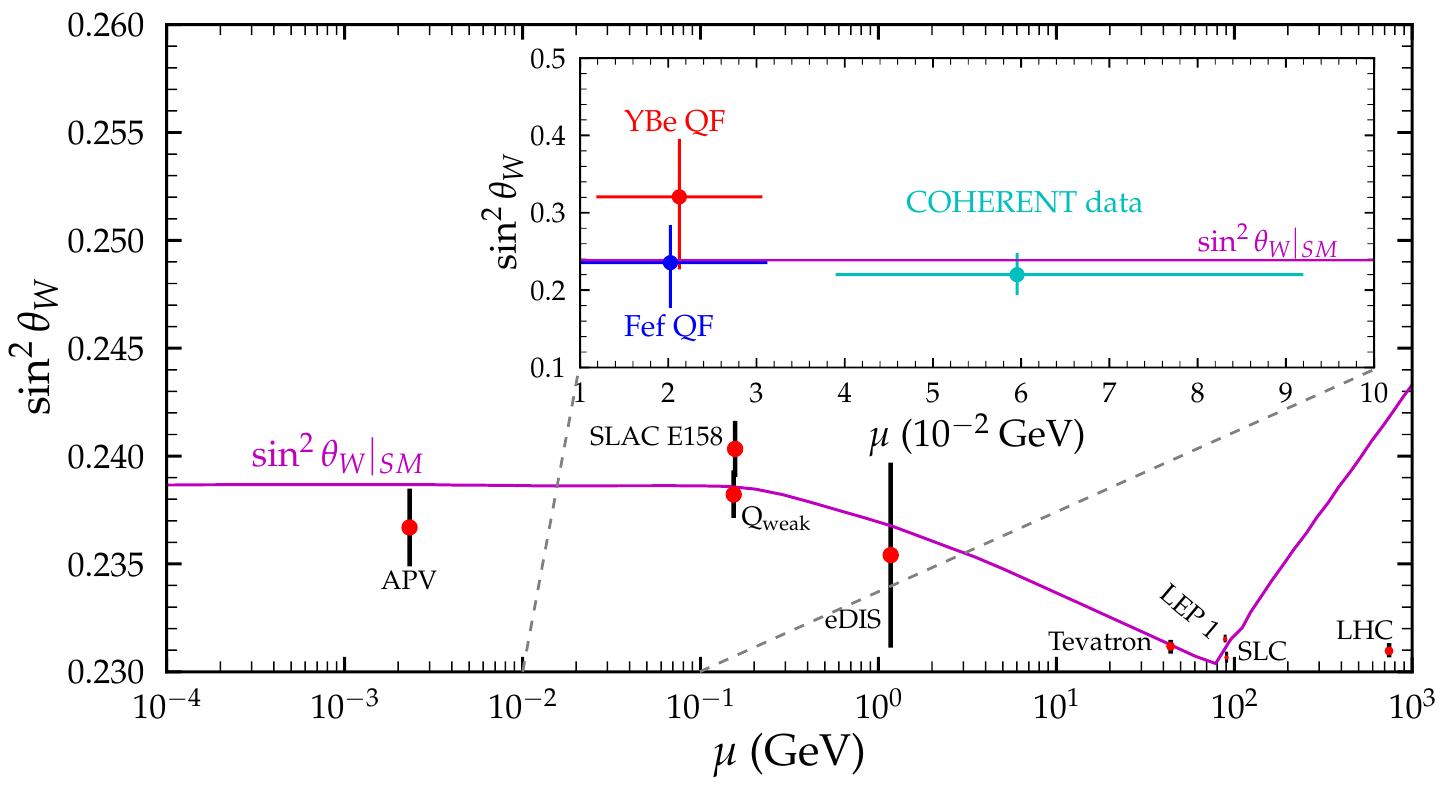}
\end{center}
\caption{RGE running of the weak mixing angle and experimental measurements across different energies.
  The inset plot shows our current result obtained from the analysis of Dresden-II data for the different QF models. A comparison with the COHERENT data~\cite{COHERENT:2021xmm} is also shown.}
\label{fig:Wienberg_Angle_Comparison_Inset_Plot}
\end{figure}
%--------------------------------------------------------------------

\subsection{Statistical analysis}
\label{subsec:Stat_Analysis}

The statistical analysis in our present work is based on the Gaussian $\chi^2$ function~\cite{Coloma:2022avw}  
\begin{equation}
\label{equn:chi_square}
\chi^2(\overrightarrow{S},a,\upbeta)=\sum_{i=1}^{130}{\left[\frac{(1+a)R^i_\text{CE$\nu$NS}(\overrightarrow{S})+R^i_\text{bkg}(\mathbf{\upbeta})-R^i_\text{exp}}{\sigma^i_\text{exp}}\right]}^2+{\left(\frac{a}{\sigma_a}\right)}^2+{\left(\frac{\beta_{M/L_1}-0.16}{\sigma_{\beta_{M/L_1}}}\right)}^2\, ,
\end{equation}
where $R^i_\text{CE$\nu$NS}$ stands for the theoretically estimated \cevns events in the $i$th bin, and $\overrightarrow{S}$ denotes the set of new physics parameters,
while $R^i_\text{exp}$ and $\sigma^i_\text{exp}$ are the experimental number of events and the corresponding uncertainty in the $i$th bin, all taken from data release~\cite{Colaresi:2022obx}.  
The neutrino flux normalization uncertainty is taken into consideration through the nuisance parameter $a$ with $\sigma_a=2$\%. Following Ref.~\cite{Colaresi:2022obx}, the uncertainty of $\beta_{M/L_1}$ is taken to be $\sigma_{\beta_{M/L_1}}=0.03$, and the prior $0.16$ is assigned.
In what follows, for a given parameter of interest from the set $\overrightarrow{S}$, our analysis involves minimization of the $\chi^2$ function over each component of $\upbeta$ and $a$.  

As a first step, we simulate the event spectra expected at Dresden-II, a procedure which has also served as a validation of our calculation. 
Figure~\ref{fig:Events} depicts the efficiency-corrected event rates within and beyond the SM for the different new physics scenarios of interest here.  
In each panel, we show both the ``\cevns only'' signal and ``\cevns + fitted background'' signal, while the event rates are zoomed in on the recoil energy range $[0.2,0.5]~\text{keV}_{ee}$,
where \cevns has a significant contribution (for the full spectra in the ROI, see Fig.~\ref{fig:SM_Events_Fitting} in the Appendix~\ref{sec:appendix}). 
Assuming that the experimental data are populated by backgrounds only, we find the best-fit value $\chi^2_\text{min} = 110.69/\text{d.o.f.}$
(d.o.f. stands for the total number of degrees of freedom).
On the other hand, by including the predicted SM \cevns signal in our analysis, we find the best-fit values
$\chi^2_\text{min} = \{ 102.66, 105.95 \}/\text{d.o.f.} $ for \{Fef, YBe\} QF---i.e., a clear improvement with respect to the background-only hypothesis.

Based on the $\chi^2$ function given in Eq.~(\ref{equn:chi_square}) and the Dresden-II data release, we proceed by presenting our results---namely, the parameter determination sensitivities within and beyond the SM for the various physics scenarios described in Sec.~\ref{sec:Theory}. 

We begin by discussing our new restrictions imposed on the SM weak mixing angle.
The corresponding $\Delta \chi^2$ profiles of  $\sin^2{\theta_W}$ for the Fef and YBe QFs are shown in Fig.~\ref{fig:Wienberg_Angle}.
The extracted 1$\sigma$ determinations of the weak mixing angle for our two QFs read 
\begin{equation*}
\begin{split}
\text{Photoneutron QF:} &\text{ }\text{ } \sin^2{\theta_W}=0.321^{+0.075}_{-0.094} \, ,\\
\text{Iron filter QF:} &\text{ }\text{ } \sin^2{\theta_W}=0.236^{+0.048}_{-0.058} \, .
\end{split}
\end{equation*}
One sees that there is a strong dependence on the QF model used.
Moreover,  our results are in good agreement with those of Ref.~\cite{AtzoriCorona:2022qrf}, though our best-fit point lies closer to the RGE prediction, especially for the case of Fef QF. 
We can also compare our results with those obtained in Refs.~\cite{AristizabalSierra:2022axl,Khan:2022jnd}, based on the background-subtracted data in~Ref.~\cite{Colaresi:2022obx}.
Exploiting the full Dresden-II data in the present analysis, we find a better agreement with the RGE prediction, but with a higher uncertainty compared to the latter studies.
In Fig.~\ref{fig:Wienberg_Angle_Comparison_Inset_Plot}, we compare our results extracted from the analysis of Dresden-II data with determinations from other probes across a wide range of energies.
Let us finally emphasize that even though the Dresden-II data constrain the weak mixing angle with lower significance than the measurement reported by the COHERENT Collaboration~\cite{COHERENT:2021xmm}, it is  interesting to notice the complementarity between the two measurements.
Clearly, the Dresden-II data fill a gap in the regime of low momentum transfer---e.g., within $10^{-2}-10^{-1}~\mathrm{GeV}$.  \\[-.5cm]

\begin{figure}[h]
\begin{center}
\includegraphics[width=0.49\textwidth]{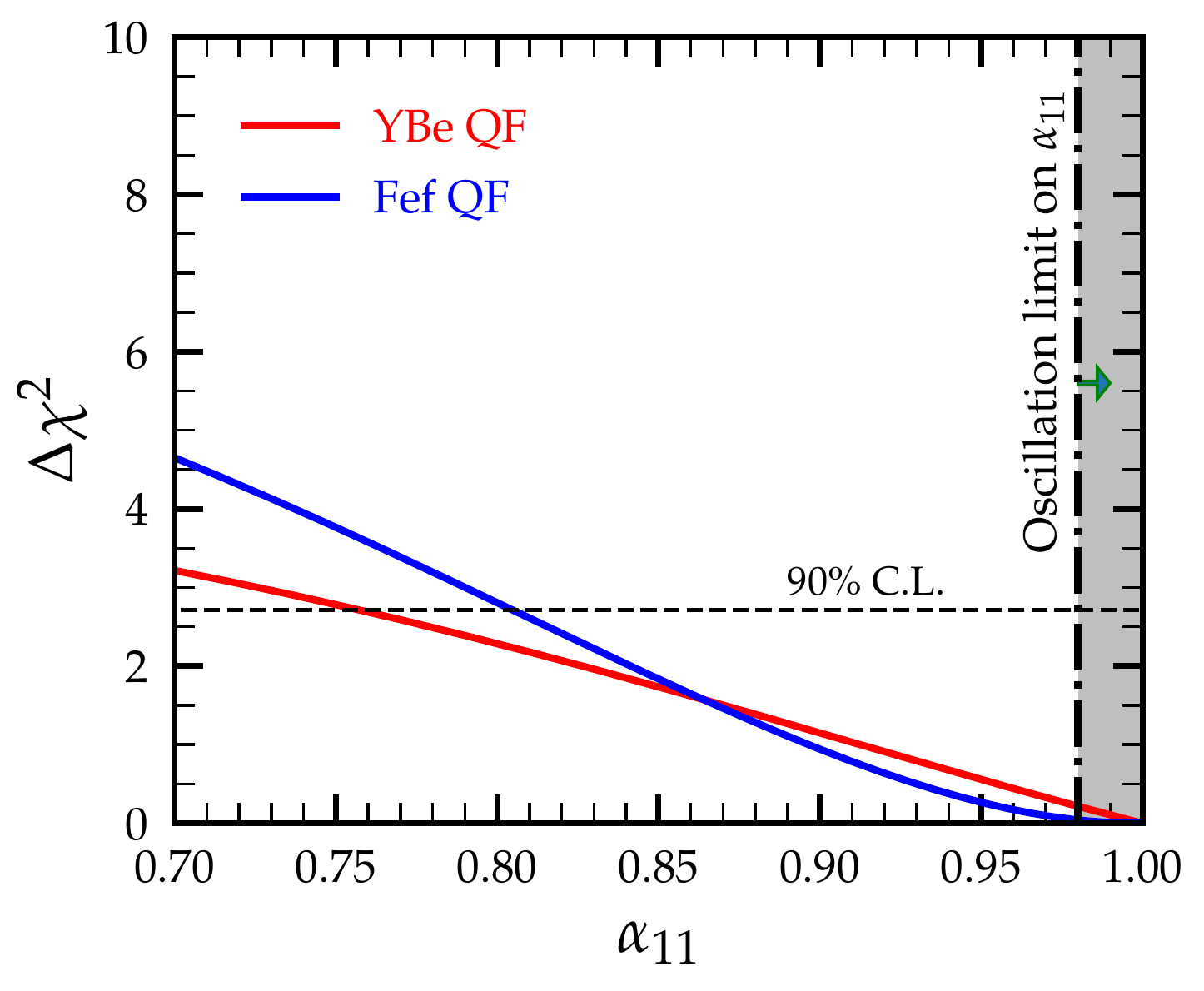}
\includegraphics[width=0.49\textwidth]{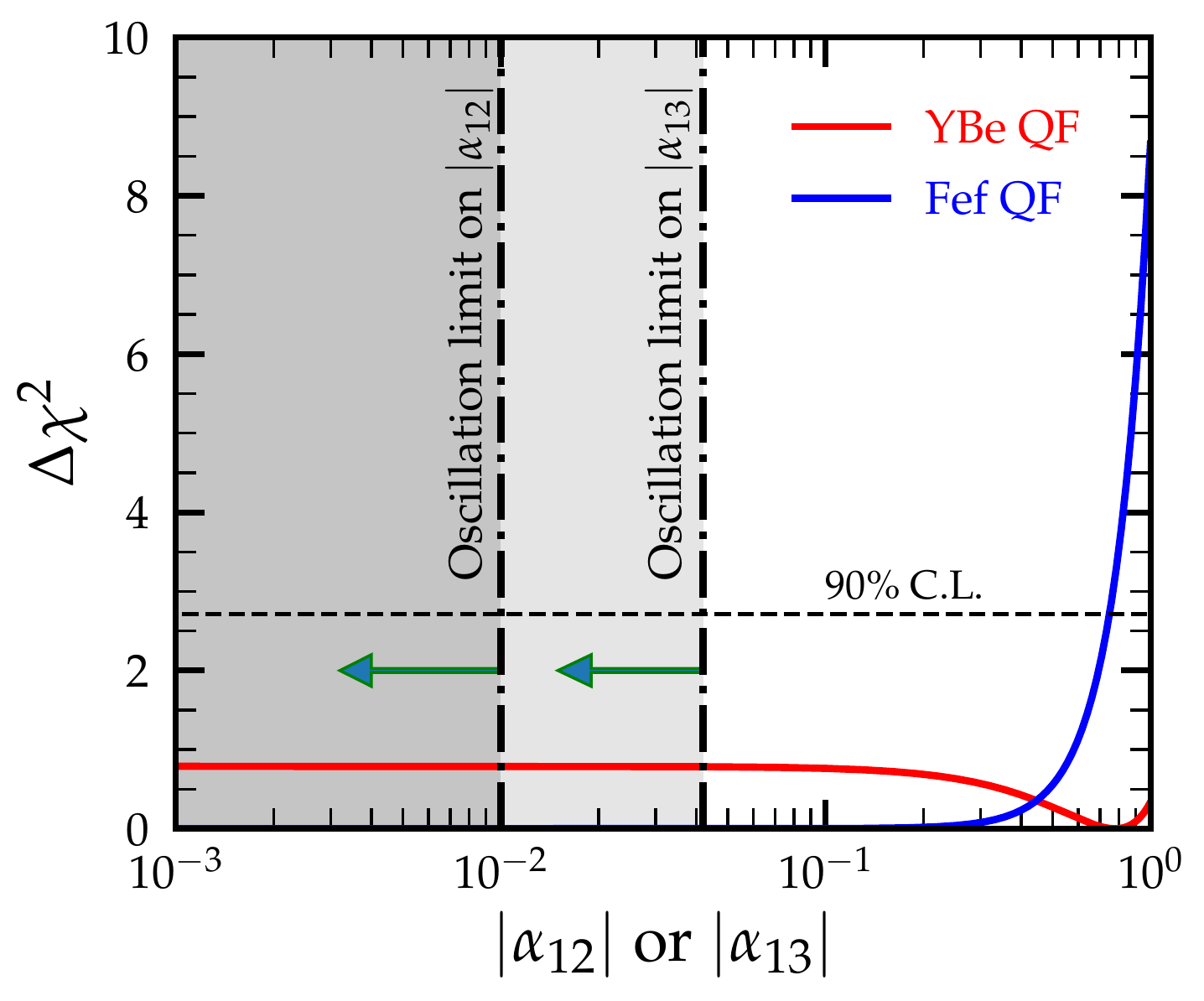}
\end{center}
\caption{\label{fig:NU} Sensitivity of Dresden-II on NU parameters, $\alpha_{11}$ (left) and $|\alpha_{12}|$ or $|\alpha_{13}|$ (right), for the Fef and YBe QF. Also shown are the corresponding  90\%~C.L. constraints extracted from  neutrino oscillation global fits~\cite{Escrihuela:2016ube}.}
\end{figure}

We now examine the possibility of using the Dresden-II data to extract information on unitarity-violation effects in the neutrino mixing matrix
in Eqs.~(\ref{equn:generalized_charged_current_weak_interaction_mixing_matrix}) and (\ref{equn:New_Physics_Matrix}). 
In Fig.~\ref{fig:NU} we present the $\Delta\chi^2$ profiles of $\alpha_{11}$ and $|\alpha_{12}|$ or $|\alpha_{13}|$, by varying one parameter at a time. 
As can be seen from Eq.(\ref{eq:events_NU}), the expected number of events has the same dependence on the off-diagonal NU parameters $|\alpha_{12}|$ and $|\alpha_{13}|$. 
As a result, they can be constrained by the Dresden-II data with exactly the same sensitivity.
We present our results for the two QF models, Fef and YBe.
As expected, the Fef-driven results are more constraining, since the pure SM event rate evaluated with the YBe QF is more suppressed, compared to the Fef spectrum.
For the case of YBe QF, the predicted event rate contains much fewer events than the spectrum calculated with the Fef QF (see Fig.~\ref{fig:Events}), and hence the NU parameters
  $|\alpha_{21}|$, $|\alpha_{31}|$ would reach values larger than 1 in order to describe the Dresden-II data, i.e. they would become unphysical.

The  90\% C.L. sensitivities on the NU parameters of interest using Dresden-II data are summarized in Table~\ref{table:NU_Parameter}.
They are compared with those derived from neutrino oscillation global fits~\cite{Escrihuela:2016ube}.
As expected, due to low statistics, large uncertainties on the background model of Dresden-II, as well as poor QF determination,
the sensitivities extracted in this work are not competitive with those achievable at the large-scale oscillation experiments.  
We note, however, that the upcoming sensitivities at the next-generation \cevns experiments are rather promising,
and they might become competitive with oscillation experiments---see, e.g., Ref.~\cite{Miranda:2020syh}.
\begin{table}[t]
\begin{adjustbox}{width=0.6\columnwidth,center}
\centering
\begin{tabular}{|c|c|c|c|}
\hline
NU parameter & Oscillations~\cite{Escrihuela:2016ube} & \hspace{3mm}YBe QF\hspace{3mm} & \hspace{3mm}Fef QF\hspace{3mm} \\
\hline
$\alpha_{11}$ & $>0.98$ & $>0.758$& $>0.805$\\
\hline
$\left|\alpha_{12}\right|$ & $<10^{-2}$ & -- & $<0.747$\\
\hline
$\left|\alpha_{13}\right|$ & $<4.2\times 10^{-2}$ & -- & $<0.747$\\
\hline
\end{tabular}
\end{adjustbox}
\caption{90\% C.L. sensitivities on NU parameters from the Dresden-II data for YBe and Fef QF. We also give, for comparison, those derived from neutrino oscillation data in Ref.~\cite{Escrihuela:2016ube}.}
\label{table:NU_Parameter}
\end{table}
\begin{figure}[t]
\begin{center}
\includegraphics[width=0.49\textwidth]{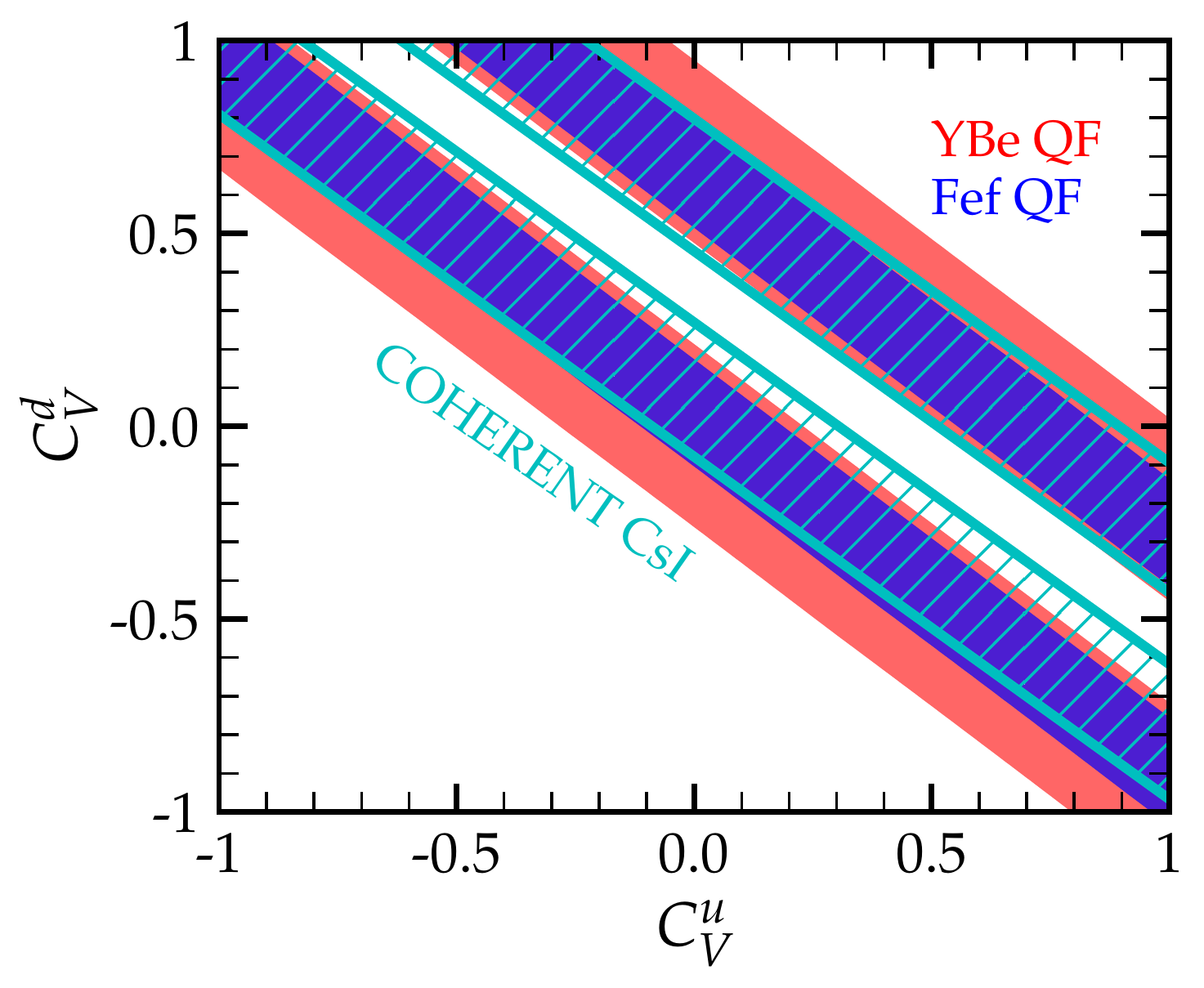}\\
\includegraphics[width=0.49\textwidth]{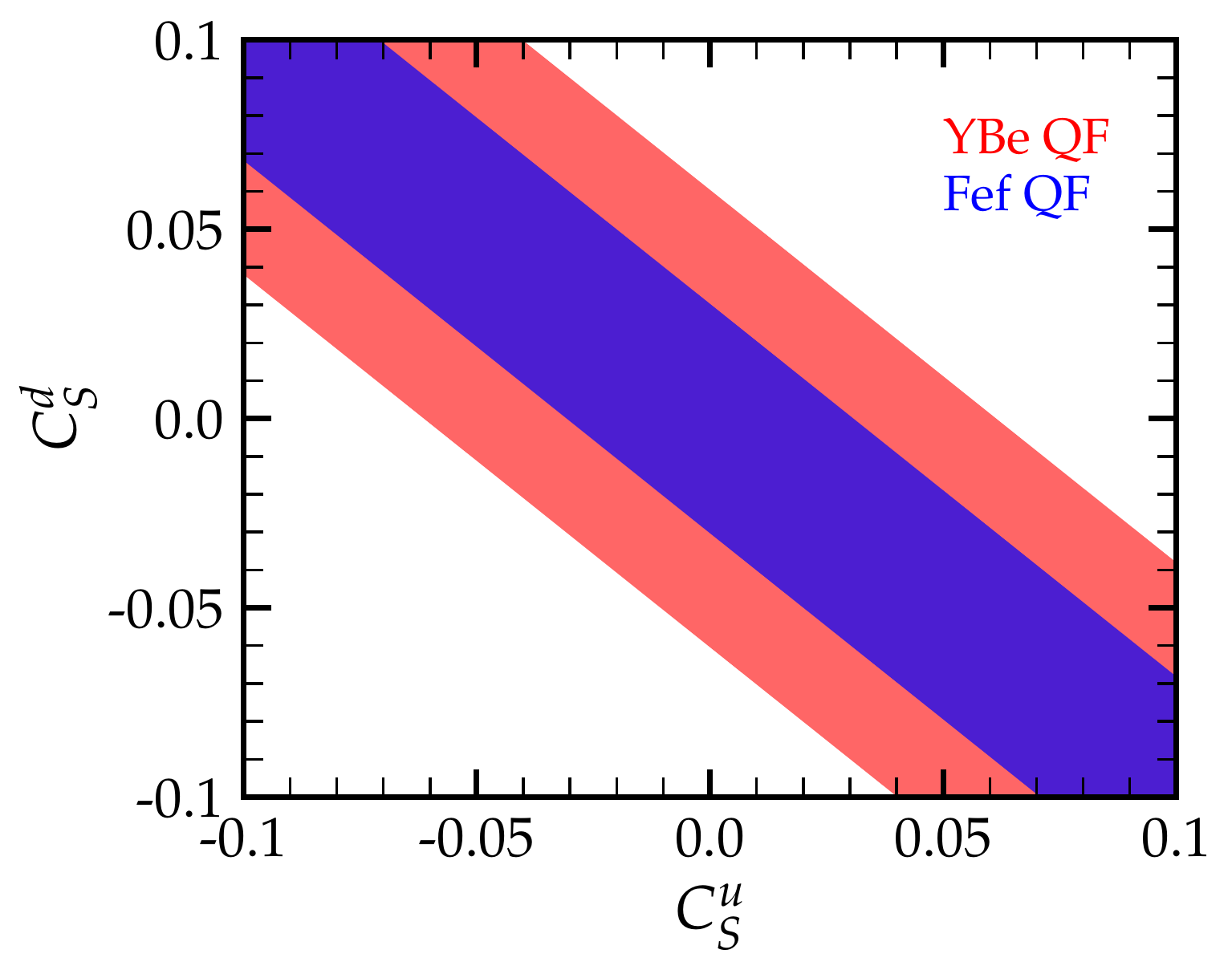}
\includegraphics[width=0.49\textwidth]{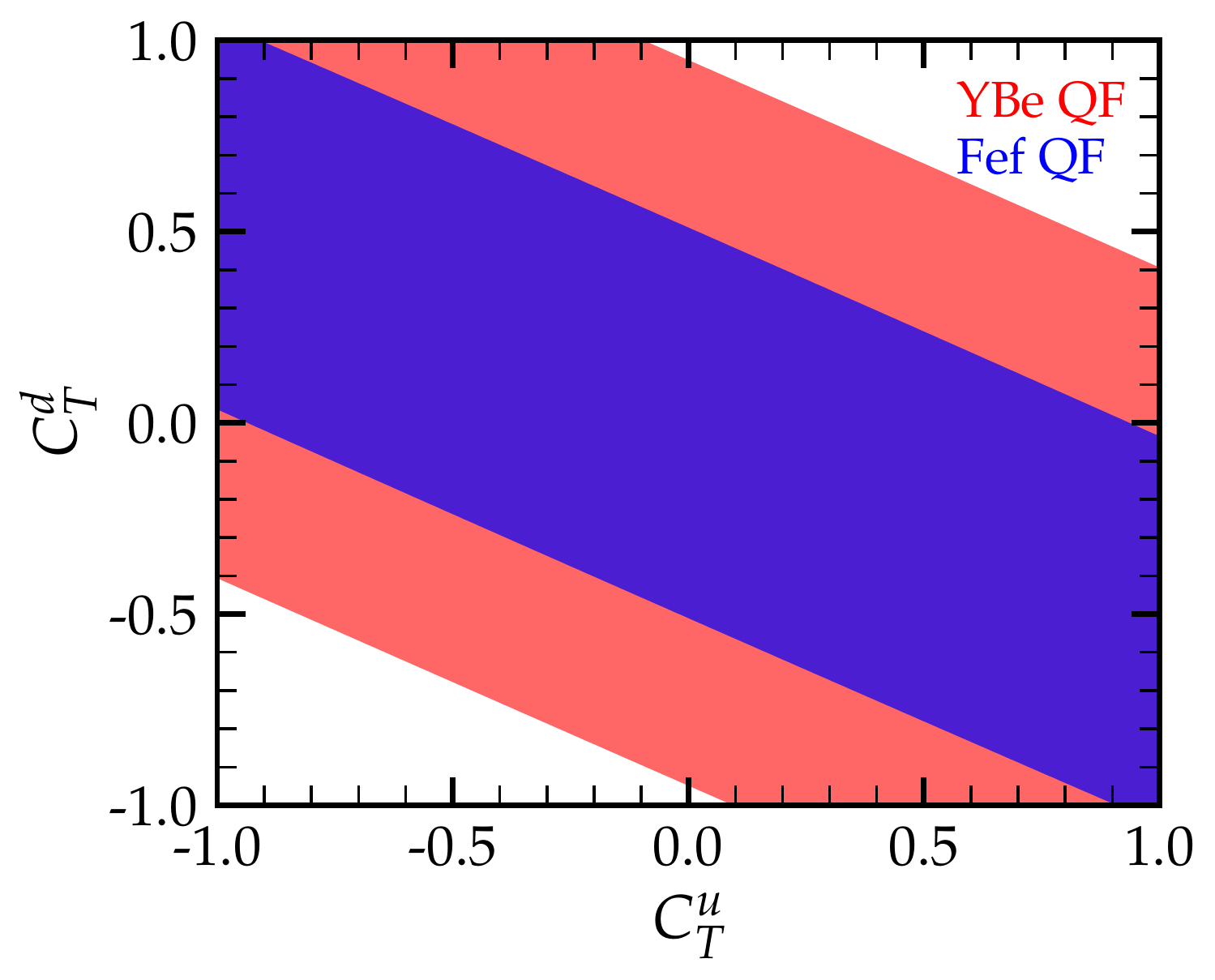}
\end{center}
\caption{90\%~C.L. allowed regions in the vector (top), scalar (lower-left) and tensor (lower-right) NGI parameters ($C_X^u, C_X^d$). We assume one nonvanishing interaction $X=S,V,T$ at a time and nonuniversal couplings for $u$ and $d$ quarks, for the Fef (in blue) and YBe (in red) quenching factors. For vector interactions, the COHERENT results~\cite{COHERENT:2021xmm} are superimposed for comparison.}
\label{fig:SVT}
\end{figure}
\begin{figure}[t]
\begin{center}
\includegraphics[width=0.49\textwidth]{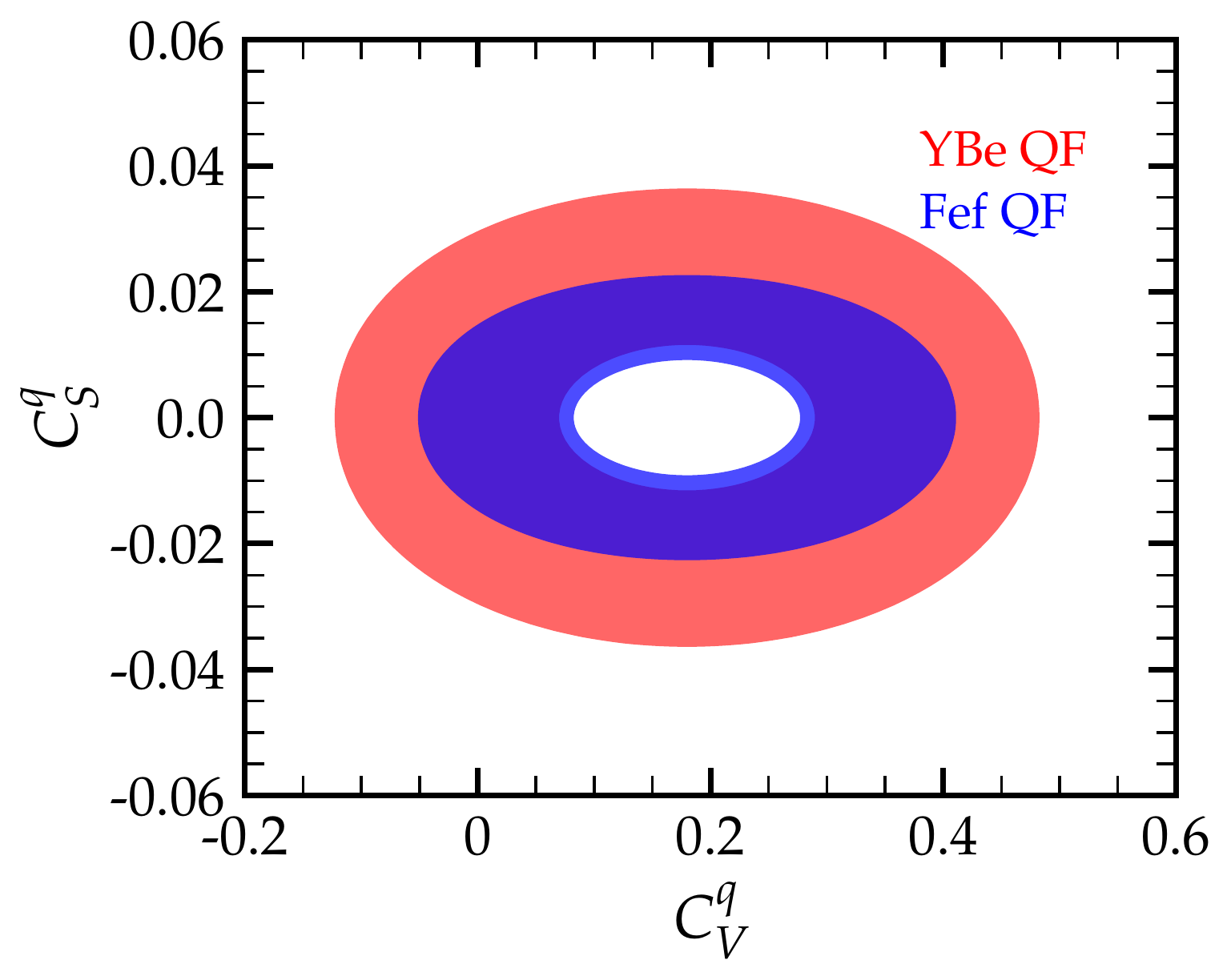}
\includegraphics[width=0.49\textwidth]{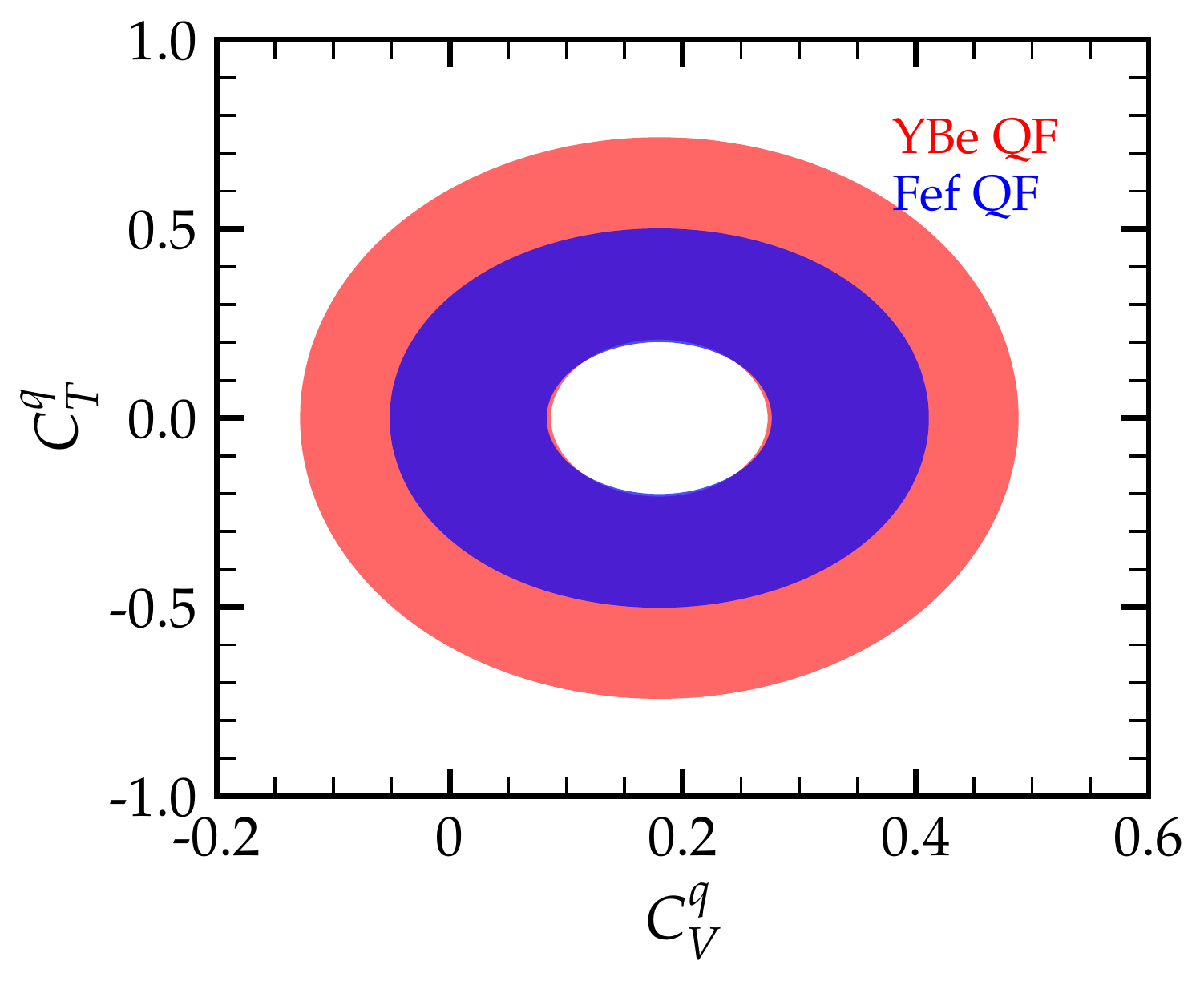}
\includegraphics[width=0.49\textwidth]{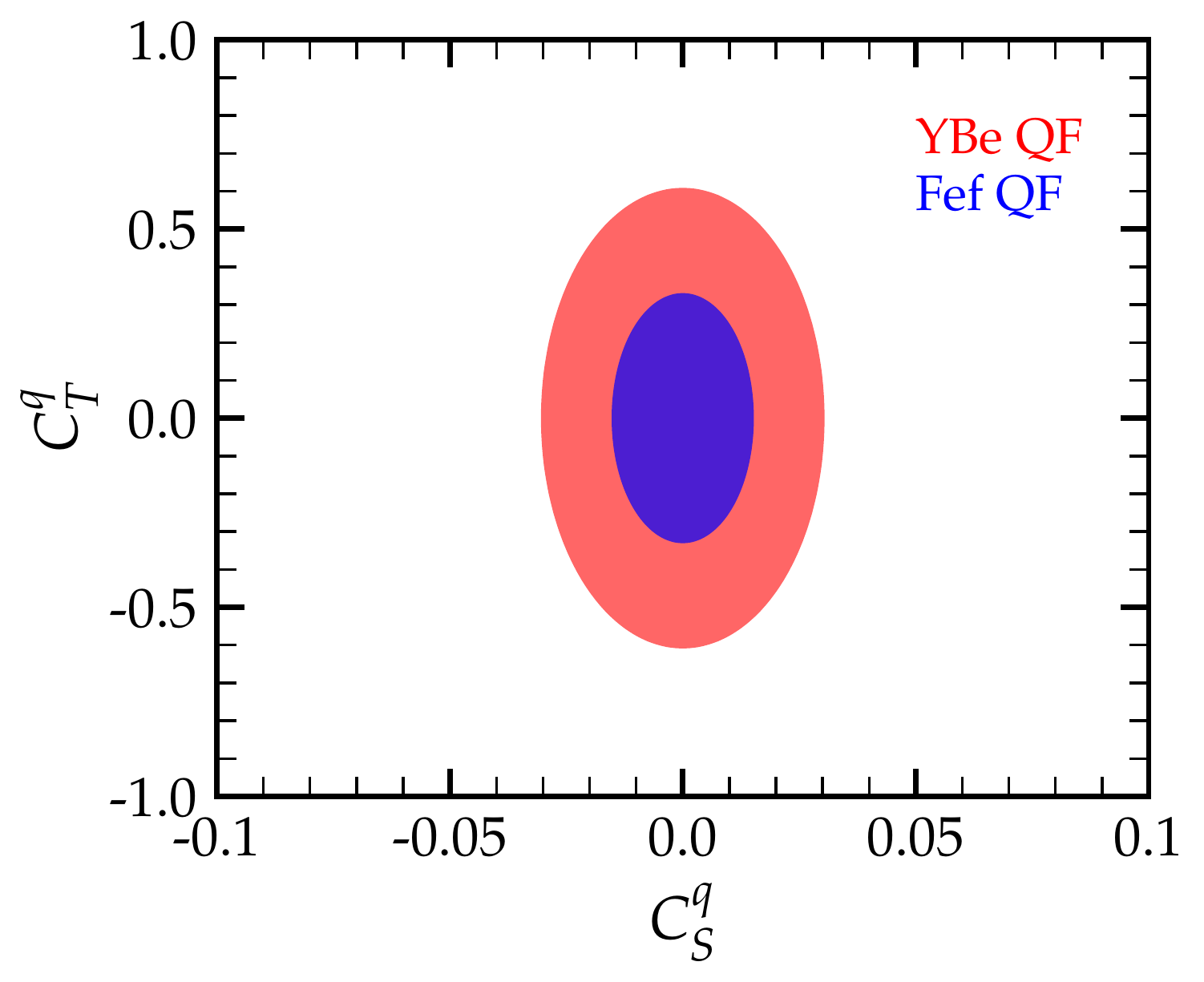}
\end{center}
\caption{Same as Fig.~\ref{fig:SVT}, but for two nonvanishing interactions $X=S,V,T$ at a time, using universal quark couplings.}
\label{fig:parameter_space}
\end{figure}

Now, we finally come to our discussion of the results of our phenomenological analysis of NGIs that includes, as a special case, the conventional NSIs.
As discussed in Sec.~\ref{subsec:NSI_xsec}, these are expected in low-scale models of neutrino mass generation; see Ref.~\cite{Boucenna:2014zba} for a more extended discussion. 
If NGIs exist, the Dresden-II reactor experiment will be sensitive to a particular combination of the $C_X^q$ coefficients, specified in Eq.~\eqref{eq:BSM_charges_cevns}.
In Fig.~\ref{fig:SVT}, we show the corresponding $90\%$ C.L. allowed regions for the NGI coefficients in the $(C_X^u, C_X^d$) plane.
These are obtained by switching on a single interaction $X$ on top of the SM, taking the other two to vanish. 
As previously, the results are given for both QFs, with the Fef constraints more stringent than the YBe ones.
Concerning the NSI case of the vector interaction, our ($C_V^u, C_V^d$) sensitivity contours (see top graph in Fig.~\ref{fig:SVT}) agree very well with those in Refs.~\cite{Coloma:2022avw,Denton:2022nol}, and with existing constraints from COHERENT~\cite{COHERENT:2021xmm}.
However, the NGI framework allows us to explore the sensitivity of Dresden-II data on scalar and tensor interactions as well~\cite{Lindner:2016wff}.  
The corresponding results are given in the lower-left (lower-right) graph of Fig.~\ref{fig:SVT} for scalar (tensor) interactions. 
As expected, in both cases, there is no destructive interference and the sensitivity contours appear as a single band, unlike the vector case
\footnote{For vector interaction, the new physics part interferes either destructively or constructively with the SM, 
  leading to the white region between the two colored stripes.}. 
Interestingly,  by comparing the three contours, one can see that the least (most) constrained NGI parameter is the vector (scalar) one. 
Indeed,  we find the best-fit values: $\chi^2_\text{min} = \{104.53, 105.16, 104.82\}/\text{d.o.f.}$ using YBe QF and $\{102.59, 102.63, 102.62\}/\text{d.o.f.}$ using Fef QF
for the $(C_S^u, C_S^d)$, $(C_V^u, C_V^d)$, $(C_T^u, C_T^d)$---i.e., the presence of NGIs implies a slight improvement in the fit with respect to the SM result, of no statistical significance.

Within the NGI framework, in principle, all interactions $X=S,V,T$ can be simultaneously active with a total number of six free NGI parameters.  
Notice that the different kinematic terms corresponding to $X=S,V,T$ interactions in the NGI cross section of Eq.(\ref{equn:NGI_xSec}) affect the shape of the \cevns event rate distribution. 
In order to reduce the degrees of freedom, while still being able to come out with intuitive results, we will make further reasonable simplifications.
First, we assume universal couplings for $u$ and $d$ quarks---i.e., $C_X^u=C_X^d=C_X^q$---thus reducing the degrees of freedom by half without losing information on the spectral shape of the signal. 
Second, we allow two different interactions to float simultaneously, while the third is set to zero.
This way, we can perform a more general analysis, with two different interaction types taken into account simultaneously in our fit.  
Figure~\ref{fig:parameter_space} shows the corresponding 90\%~C.L. sensitivity contours in the $(C_V^{q}, C_S^{q})$, $(C_V^{q}, C_T^{q})$, and $(C_S^{q}, C_T^{q})$ planes for the two QFs.
One sees that, in the presence of the vector NGI there is a destructive interference, leaving a ``hole'' in the $(C_V^{q}, C_S^{q})$, $(C_V^{q}, C_T^{q})$ planes.  
The best-fit values are:
$\chi^2_\text{min} = \{103.81, 104.39, 104.54\}/\text{d.o.f.}$ using YBe QF, and $\{102.01, 102.05, 102.59\}/\text{d.o.f.}$ using Fef QF for $(C_V^{q}, C_S^{q})$, $(C_V^{q}, C_T^{q})$, and $(C_S^{q}, C_T^{q})$, respectively.
Thus, we conclude that a combined analysis with two nonzero NGIs at a time leads to a slight improvement in the fit compared to the pure SM case, or a single nonvanishing NGI type.

\section{Conclusions}
\label{sec:conclusions}

We have analyzed the new data released by the Dresden-II Collaboration which reported, for the first time, a suggestive evidence of \cevns observation using reactor antineutrinos. 
We performed a careful statistical analysis of the event spectral distributions, estimating the expected background sources according to the prescription given by the Dresden-II Collaboration: see Fig.~\ref{fig:Events}.
Using the available data, we explored different physics schemes within and beyond the SM. 
We first examined the new determination of the weak mixing angle in the low-energy regime, given in Fig.~\ref{fig:Wienberg_Angle}. 
Although our Dresden-II determination is not as competitive as the existing one from the COHERENT data, the best-fit value is in better agreement with the theoretical
prediction from RGE extrapolations (Fig.~\ref{fig:Wienberg_Angle_Comparison_Inset_Plot}).
However, one should note that the best-fit value differs dramatically for the two QF models we have adopted.

Concerning new physics scenarios, we focused on two different examples namely, the potential violation of unitarity in the neutrino mixing matrix,
as well as the presence of new neutrino interactions due to heavy mediators. 
As seen in Fig.~\ref{fig:NU}, current Dresden-II data do not have sufficient statistics to place a competitive constraint on unitarity violation parameters.
However, short baseline reactor \cevns experiments may provide a promising unitarity violation probe in the long run.  
Finally, we examined the Dresden-II limits on the most general neutrino interactions due to heavy mediators, given in Figs.~\ref{fig:SVT} and \ref{fig:parameter_space}. 
These are competitive with those obtained at COHERENT for the case of vector interaction. We have also obtained the corresponding sensitivities for scalar and tensor \cevns interactions.
Though not significant, their interplay may improve the description of the Dresden-II data.

\acknowledgments

\begin{justify}
We thank Prof. Juan I. Collar for providing us the updated version of Ref.~\cite{Colaresi:2022obx} containing ranges of the free background parameters for the Dresden-II analysis
as well as their $1~\sigma$ limit.
This work has been supported by the SERB, Government of India grant SRG/2020/002303,
by the Spanish Agencia Estatal de Investigacion under grant N. PID2020-113775GB-I00 (AEI/10.13039/501100011033) and
by Generalitat Valenciana under Prometeo grant N. CIPROM/2021/054.
\end{justify}

\appendix 
\section{Details of the data fitting}
\label{sec:appendix}

\begin{figure}[t]
\begin{center}
\includegraphics[width=0.49\textwidth]{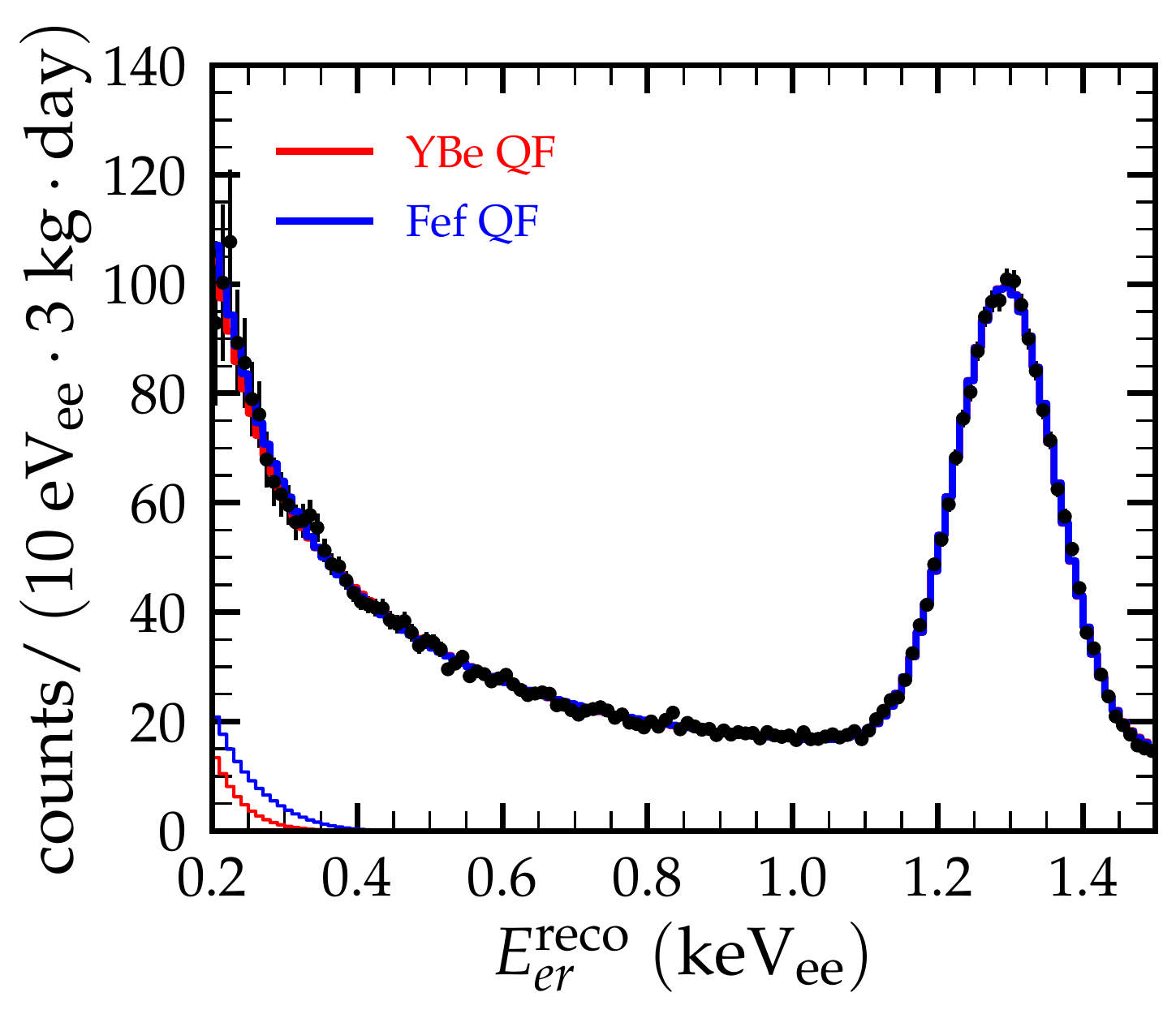}
\end{center}
\caption{SM ``\cevns only'' (thin lines) and ``\cevns + fitted background'' (thick lines) signal in the full ROI of the Dresden-II measurement.}
\label{fig:SM_Events_Fitting}
\end{figure}

In this section, we briefly discuss the details of the determination of the expected background events for the Dresden-II data used in this work.  
As already mentioned in Sec.~\ref{sec:results}, the background model characterized by seven free parameters is reported in Ref.~\cite{Colaresi:2022obx}.
Note also that every analysis requires simultaneous fitting of the free background parameters in addition to the physics parameters of interest within or beyond the SM.
Hence, in this work we paid special attention in adopting a careful fitting procedure.  
In Fig.~\ref{fig:SM_Events_Fitting}, we  demonstrate the SM ``\cevns only'' signal and the ``\cevns + fitted background''  signal in the full ROI of the Dresden-II experiment. 
In contrast to Refs.~\cite{Coloma:2022avw, AtzoriCorona:2022qrf}, which employed a differential spectrum, here we make different background assumptions.
Namely, the background model of Eq.(\ref{equn:bkg}) is taken to be an integrated spectrum for which the parameters $N_\text{epith}$, $A_\text{epith}$, and $A_{L_1, L_2, L_M}$ are given in units of
$\mathrm{counts/\left (10~eV_{ee} \cdot 3~kg \cdot day \right)}$, in line with the approach followed by the Dresden-II Collaboration.
Although this may not lead to significant differences, we stress that in our analysis the background parameters are varied within their allowed ranges, unspecified in Refs.~\cite{Coloma:2022avw, AtzoriCorona:2022qrf}.  
The most notable difference with those works is that here the exponential decay function characterizing the epithermal neutron background component is taken to be centered at $E_\text{epith}=0.2~\mathrm{keV_{ee}}$ instead of zero.
The allowed ranges and the best-fit values of free background parameters are listed in Table~\ref{table:Bkg_Parameters_Best_Fit_Values}. 
For each case, the best-fit values we obtain are in agreement with the corresponding $1~\sigma$ limit reported in the ancillary file of Ref.~\cite{Colaresi:2022obx}.

\begin{table}[t]
\centering
\begin{tabular}{|c|c|c|c|}
\hline
Background Parameters ($\upbeta$)  & \hspace{3mm}Allowed range\hspace{3mm} & \hspace{3mm}YBe QF\hspace{3mm} & \hspace{3mm}Fef QF\hspace{3mm} \\
\hline
$N_\text{epith}$ & $[0,~25]$ & $14.082$& $13.779$\\
$A_\text{epith}$  & $[0,~150]$ & $66.505$ & $62.947$\\
$\tau_\text{epith}\text{ (keV)}$ & $[0,~2]$ & $0.249$ & $0.262$\\
$A_{L_1}$ & $[70,~250]$ & $84.577$ & $84.63$\\
$E_{L_1}\text{ (keV)}$ & $[1.2,~1.4]$ & $1.29228$ & $1.29237$\\
$\sigma_{L_1}\text{ (keV)}$ & $[0.04,~0.1]$ & $0.0694$ & $0.06963$\\
$\beta_{M/L_1}$ & $[0,~0.3]$ & $0.17$ & $0.158$\\
\hline
Fitted events (SM $+$ Background)&&$5080.29$&$5103.4$\\
\hline
$\chi^2_\text{min}$ & & $105.95/\text{d.o.f.}$ & $102.66/\text{d.o.f.}$\\
\hline
\end{tabular}
\caption{Best-fit values for the background model parameters and total number of signal events. The values of $N_\text{epith}$, $A_\text{epith}$, $A_{L_1}$, as well as the fitted events are all given in units of $\mathrm{[counts/\left (10~eV_{ee} \cdot 3~kg \cdot day \right)]}$.}
\label{table:Bkg_Parameters_Best_Fit_Values}
\end{table}

\newpage
\bibliographystyle{utphys}
\bibliography{bibliography}

\providecommand{\href}[2]{#2}\begingroup\raggedright\begin{thebibliography}{10}

\bibitem{Freedman:1973yd}
D.~Z. Freedman, ``{Coherent Neutrino Nucleus Scattering as a Probe of the Weak
  Neutral Current},'' \href{http://dx.doi.org/10.1103/PhysRevD.9.1389}{{\em
  Phys. Rev. D} {\bfseries 9} (1974) 1389--1392}.

\bibitem{COHERENT:2017ipa}
{\bfseries COHERENT} Collaboration, D.~Akimov {\em et~al.}, ``{Observation of
  Coherent Elastic Neutrino-Nucleus Scattering},''
  \href{http://dx.doi.org/10.1126/science.aao0990}{{\em Science} {\bfseries
  357} no.~6356, (2017) 1123--1126},
  \href{http://arxiv.org/abs/1708.01294}{{\ttfamily arXiv:1708.01294
  [nucl-ex]}}.

\bibitem{COHERENT:2021xmm}
{\bfseries COHERENT} Collaboration, D.~Akimov {\em et~al.}, ``{Measurement of
  the Coherent Elastic Neutrino-Nucleus Scattering Cross Section on CsI by
  COHERENT},'' \href{http://dx.doi.org/10.1103/PhysRevLett.129.081801}{{\em
  Phys. Rev. Lett.} {\bfseries 129} no.~8, (2022) 081801},
  \href{http://arxiv.org/abs/2110.07730}{{\ttfamily arXiv:2110.07730
  [hep-ex]}}.

\bibitem{COHERENT:2020iec}
{\bfseries COHERENT} Collaboration, D.~Akimov {\em et~al.}, ``{First
  Measurement of Coherent Elastic Neutrino-Nucleus Scattering on Argon},''
  \href{http://dx.doi.org/10.1103/PhysRevLett.126.012002}{{\em Phys. Rev.
  Lett.} {\bfseries 126} no.~1, (2021) 012002},
  \href{http://arxiv.org/abs/2003.10630}{{\ttfamily arXiv:2003.10630
  [nucl-ex]}}.

\bibitem{CONNIE:2021ggh}
{\bfseries CONNIE} Collaboration, A.~Aguilar-Arevalo {\em et~al.}, ``{Search
  for coherent elastic neutrino-nucleus scattering at a nuclear reactor with
  CONNIE 2019 data},'' \href{http://dx.doi.org/10.1007/JHEP05(2022)017}{{\em
  JHEP} {\bfseries 05} (2022) 017},
  \href{http://arxiv.org/abs/2110.13033}{{\ttfamily arXiv:2110.13033
  [hep-ex]}}.

\bibitem{CONUS:2020skt}
{\bfseries CONUS} Collaboration, H.~Bonet {\em et~al.}, ``{Constraints on
  Elastic Neutrino Nucleus Scattering in the Fully Coherent Regime from the
  CONUS Experiment},''
  \href{http://dx.doi.org/10.1103/PhysRevLett.126.041804}{{\em Phys. Rev.
  Lett.} {\bfseries 126} no.~4, (2021) 041804},
  \href{http://arxiv.org/abs/2011.00210}{{\ttfamily arXiv:2011.00210
  [hep-ex]}}.

\bibitem{CONUS:2021dwh}
{\bfseries CONUS} Collaboration, H.~Bonet {\em et~al.}, ``{Novel constraints on
  neutrino physics beyond the standard model from the CONUS experiment},''
  \href{http://dx.doi.org/10.1007/JHEP05(2022)085}{{\em JHEP} {\bfseries 05}
  (2022) 085}, \href{http://arxiv.org/abs/2110.02174}{{\ttfamily
  arXiv:2110.02174 [hep-ph]}}.

\bibitem{nuGeN:2022bmg}
{\bfseries \ensuremath{\nu}GeN} Collaboration, I.~Alekseev {\em et~al.},
  ``{First results of the $\ensuremath{\nu}\mathrm{GeN}$ experiment on coherent
  elastic neutrino-nucleus scattering},''
  \href{http://dx.doi.org/10.1103/PhysRevD.106.L051101}{{\em Phys. Rev. D}
  {\bfseries 106} no.~5, (2022) L051101},
  \href{http://arxiv.org/abs/2205.04305}{{\ttfamily arXiv:2205.04305
  [nucl-ex]}}.

\bibitem{MINER:2016igy}
{\bfseries MINER} Collaboration, G.~Agnolet {\em et~al.}, ``{Background Studies
  for the MINER Coherent Neutrino Scattering Reactor Experiment},''
  \href{http://dx.doi.org/10.1016/j.nima.2017.02.024}{{\em Nucl. Instrum. Meth.
  A} {\bfseries 853} (2017) 53--60},
  \href{http://arxiv.org/abs/1609.02066}{{\ttfamily arXiv:1609.02066
  [physics.ins-det]}}.

\bibitem{Billard:2016giu}
J.~Billard {\em et~al.}, ``{Coherent Neutrino Scattering with Low Temperature
  Bolometers at Chooz Reactor Complex},''
  \href{http://dx.doi.org/10.1088/1361-6471/aa83d0}{{\em J. Phys. G} {\bfseries
  44} no.~10, (2017) 105101}, \href{http://arxiv.org/abs/1612.09035}{{\ttfamily
  arXiv:1612.09035 [physics.ins-det]}}.

\bibitem{Strauss:2017cuu}
R.~Strauss {\em et~al.}, ``{The $\nu$-cleus experiment: A gram-scale
  fiducial-volume cryogenic detector for the first detection of coherent
  neutrino-nucleus scattering},''
  \href{http://dx.doi.org/10.1140/epjc/s10052-017-5068-2}{{\em Eur. Phys. J. C}
  {\bfseries 77} (2017) 506}, \href{http://arxiv.org/abs/1704.04320}{{\ttfamily
  arXiv:1704.04320 [physics.ins-det]}}.

\bibitem{Wong:2015kgl}
H.~T.-K. Wong, ``{Taiwan EXperiment On NeutrinO \textemdash{} History and
  Prospects},'' \href{http://dx.doi.org/10.1142/S0217751X18300144}{{\em The
  Universe} {\bfseries 3} no.~4, (2015) 22--37},
  \href{http://arxiv.org/abs/1608.00306}{{\ttfamily arXiv:1608.00306
  [hep-ex]}}.

\bibitem{Fernandez-Moroni:2020yyl}
G.~Fernandez-Moroni, P.~A.~N. Machado, I.~Martinez-Soler, Y.~F. Perez-Gonzalez,
  D.~Rodrigues, and S.~Rosauro-Alcaraz, ``{The physics potential of a reactor
  neutrino experiment with Skipper CCDs: Measuring the weak mixing angle},''
  \href{http://dx.doi.org/10.1007/JHEP03(2021)186}{{\em JHEP} {\bfseries 03}
  (2021) 186}, \href{http://arxiv.org/abs/2009.10741}{{\ttfamily
  arXiv:2009.10741 [hep-ph]}}.

\bibitem{SBC:2021yal}
{\bfseries SBC, CE\ensuremath{\nu}NS Theory Group at IF-UNAM} Collaboration,
  L.~J. Flores {\em et~al.}, ``{Physics reach of a low threshold scintillating
  argon bubble chamber in coherent elastic neutrino-nucleus scattering reactor
  experiments},'' \href{http://dx.doi.org/10.1103/PhysRevD.103.L091301}{{\em
  Phys. Rev. D} {\bfseries 103} no.~9, (2021) L091301},
  \href{http://arxiv.org/abs/2101.08785}{{\ttfamily arXiv:2101.08785
  [hep-ex]}}.

\bibitem{Baxter:2019mcx}
D.~Baxter {\em et~al.}, ``{Coherent Elastic Neutrino-Nucleus Scattering at the
  European Spallation Source},''
  \href{http://dx.doi.org/10.1007/JHEP02(2020)123}{{\em JHEP} {\bfseries 02}
  (2020) 123}, \href{http://arxiv.org/abs/1911.00762}{{\ttfamily
  arXiv:1911.00762 [physics.ins-det]}}.

\bibitem{CCM:2021yzc}
{\bfseries CCM} Collaboration, A.~A. Aguilar-Arevalo {\em et~al.}, ``{First
  Leptophobic Dark Matter Search from the Coherent\textendash{}CAPTAIN-Mills
  Liquid Argon Detector},''
  \href{http://dx.doi.org/10.1103/PhysRevLett.129.021801}{{\em Phys. Rev.
  Lett.} {\bfseries 129} no.~2, (2022) 021801},
  \href{http://arxiv.org/abs/2109.14146}{{\ttfamily arXiv:2109.14146
  [hep-ex]}}.

\bibitem{Colaresi:2021kus}
J.~Colaresi, J.~I. Collar, T.~W. Hossbach, A.~R.~L. Kavner, C.~M. Lewis, A.~E.
  Robinson, and K.~M. Yocum, ``{First results from a search for coherent
  elastic neutrino-nucleus scattering at a reactor site},''
  \href{http://dx.doi.org/10.1103/PhysRevD.104.072003}{{\em Phys. Rev. D}
  {\bfseries 104} no.~7, (2021) 072003},
  \href{http://arxiv.org/abs/2108.02880}{{\ttfamily arXiv:2108.02880
  [hep-ex]}}.

\bibitem{Colaresi:2022obx}
J.~Colaresi, J.~I. Collar, T.~W. Hossbach, C.~M. Lewis, and K.~M. Yocum,
  ``{Measurement of Coherent Elastic Neutrino-Nucleus Scattering from Reactor
  Antineutrinos},''
  \href{http://dx.doi.org/10.1103/PhysRevLett.129.211802}{{\em Phys. Rev.
  Lett.} {\bfseries 129} no.~21, (2022) 211802},
  \href{http://arxiv.org/abs/2202.09672}{{\ttfamily arXiv:2202.09672
  [hep-ex]}}.

\bibitem{AristizabalSierra:2022axl}
D.~Aristizabal~Sierra, V.~De~Romeri, and D.~K. Papoulias, ``{Consequences of
  the Dresden-II reactor data for the weak mixing angle and new physics},''
  \href{http://dx.doi.org/10.1007/JHEP09(2022)076}{{\em JHEP} {\bfseries 09}
  (2022) 076}, \href{http://arxiv.org/abs/2203.02414}{{\ttfamily
  arXiv:2203.02414 [hep-ph]}}.

\bibitem{Khan:2022jnd}
A.~N. Khan, ``{$\sin^2\theta_W$ and neutrino electromagnetic interactions in
  CE$\bar{\nu}_e$NS with different quenching factors},''
  \href{http://arxiv.org/abs/2203.08892}{{\ttfamily arXiv:2203.08892
  [hep-ph]}}.

\bibitem{AtzoriCorona:2022qrf}
M.~Atzori~Corona, M.~Cadeddu, N.~Cargioli, F.~Dordei, C.~Giunti, Y.~F. Li,
  C.~A. Ternes, and Y.~Y. Zhang, ``{Impact of the Dresden-II and COHERENT
  neutrino scattering data on neutrino electromagnetic properties and
  electroweak physics},'' \href{http://dx.doi.org/10.1007/JHEP09(2022)164}{{\em
  JHEP} {\bfseries 09} (2022) 164},
  \href{http://arxiv.org/abs/2205.09484}{{\ttfamily arXiv:2205.09484
  [hep-ph]}}.

\bibitem{Liao:2022hno}
J.~Liao, H.~Liu, and D.~Marfatia, ``{Implications of the first evidence for
  coherent elastic scattering of reactor neutrinos},''
  \href{http://dx.doi.org/10.1103/PhysRevD.106.L031702}{{\em Phys. Rev. D}
  {\bfseries 106} no.~3, (2022) L031702},
  \href{http://arxiv.org/abs/2202.10622}{{\ttfamily arXiv:2202.10622
  [hep-ph]}}.

\bibitem{Coloma:2022avw}
P.~Coloma, I.~Esteban, M.~C. Gonzalez-Garcia, L.~Larizgoitia, F.~Monrabal, and
  S.~Palomares-Ruiz, ``{Bounds on new physics with data of the Dresden-II
  reactor experiment and COHERENT},''
  \href{http://dx.doi.org/10.1007/JHEP05(2022)037}{{\em JHEP} {\bfseries 05}
  (2022) 037}, \href{http://arxiv.org/abs/2202.10829}{{\ttfamily
  arXiv:2202.10829 [hep-ph]}}.

\bibitem{Denton:2022nol}
P.~B. Denton and J.~Gehrlein, ``{New constraints on the dark side of
  non-standard interactions from reactor neutrino scattering data},''
  \href{http://dx.doi.org/10.1103/PhysRevD.106.015022}{{\em Phys. Rev. D}
  {\bfseries 106} no.~1, (2022) 015022},
  \href{http://arxiv.org/abs/2204.09060}{{\ttfamily arXiv:2204.09060
  [hep-ph]}}.

\bibitem{Boucenna:2014zba}
S.~M. Boucenna, S.~Morisi, and J.~W.~F. Valle, ``{The low-scale approach to
  neutrino masses},'' \href{http://dx.doi.org/10.1155/2014/831598}{{\em
  Adv.High Energy Phys.} {\bfseries 2014} (2014) 831598},
  \href{http://arxiv.org/abs/1404.3751}{{\ttfamily arXiv:1404.3751 [hep-ph]}}.

\bibitem{Mohapatra:1986bd}
R.~N. Mohapatra and J.~W.~F. Valle, ``{Neutrino Mass and Baryon Number
  Nonconservation in Superstring Models},''
  \href{http://dx.doi.org/10.1103/PhysRevD.34.1642}{{\em Phys. Rev. D}
  {\bfseries 34} (1986) 1642}.

\bibitem{Gonzalez-Garcia:1988okv}
M.~Gonzalez-Garcia and J.~W.~F. Valle, ``{Fast Decaying Neutrinos and
  Observable Flavor Violation in a New Class of Majoron Models},''
  \href{http://dx.doi.org/10.1016/0370-2693(89)91131-3}{{\em Phys.Lett.B}
  {\bfseries 216} (1989) 360--366}.

\bibitem{Akhmedov:1995ip}
E.~K. Akhmedov {\em et~al.}, ``{Left-right symmetry breaking in NJL
  approach},'' \href{http://dx.doi.org/10.1016/0370-2693(95)01504-3}{{\em
  Phys.Lett.B} {\bfseries 368} (1996) 270--280},
  \href{http://arxiv.org/abs/hep-ph/9507275}{{\ttfamily arXiv:hep-ph/9507275
  [hep-ph]}}.

\bibitem{Akhmedov:1995vm}
E.~K. Akhmedov {\em et~al.}, ``{Dynamical left-right symmetry breaking},''
  \href{http://dx.doi.org/10.1103/PhysRevD.53.2752}{{\em Phys.Rev.D} {\bfseries
  53} (1996) 2752--2780}, \href{http://arxiv.org/abs/hep-ph/9509255}{{\ttfamily
  arXiv:hep-ph/9509255 [hep-ph]}}.

\bibitem{Malinsky:2005bi}
M.~Malinsky, J.~Romao, and J.~W.~F. Valle, ``{Novel supersymmetric SO(10)
  seesaw mechanism},''
  \href{http://dx.doi.org/10.1103/PhysRevLett.95.161801}{{\em Phys.Rev.Lett.}
  {\bfseries 95} (2005) 161801},
  \href{http://arxiv.org/abs/hep-ph/0506296}{{\ttfamily arXiv:hep-ph/0506296
  [hep-ph]}}.

\bibitem{Dittmar:1989yg}
M.~Dittmar, A.~Santamaria, M.~Gonzalez-Garcia, and J.~W.~F. Valle,
  ``{Production Mechanisms and Signatures of Isosinglet Neutral Heavy Leptons
  in $Z^0$ Decays},''
  \href{http://dx.doi.org/10.1016/0550-3213(90)90028-C}{{\em Nucl.Phys.B}
  {\bfseries 332} (1990) 1--19}.

\bibitem{Gonzalez-Garcia:1990sbd}
M.~Gonzalez-Garcia, A.~Santamaria, and J.~W.~F. Valle, ``{Isosinglet Neutral
  Heavy Lepton Production in $Z$ Decays and Neutrino Mass},''
  \href{http://dx.doi.org/10.1016/0550-3213(90)90573-V}{{\em Nucl.Phys.B}
  {\bfseries 342} (1990) 108--126}.

\bibitem{AguilarSaavedra:2012fu}
J.~Aguilar-Saavedra, F.~Deppisch, O.~Kittel, and J.~W.~F. Valle, ``{Flavour in
  heavy neutrino searches at the LHC},''
  \href{http://dx.doi.org/10.1103/PhysRevD.85.091301}{{\em Phys.Rev.D}
  {\bfseries 85} (2012) 091301},
  \href{http://arxiv.org/abs/1203.5998}{{\ttfamily arXiv:1203.5998 [hep-ph]}}.

\bibitem{Das:2012ii}
S.~Das, F.~Deppisch, O.~Kittel, and J.~W.~F. Valle, ``{Heavy Neutrinos and
  Lepton Flavour Violation in Left-Right Symmetric Models at the LHC},''
  \href{http://dx.doi.org/10.1103/PhysRevD.86.055006}{{\em Phys.Rev.D}
  {\bfseries 86} (2012) 055006},
  \href{http://arxiv.org/abs/1206.0256}{{\ttfamily arXiv:1206.0256 [hep-ph]}}.

\bibitem{Deppisch:2013cya}
F.~F. Deppisch, N.~Desai, and J.~W.~F. Valle, ``{Is charged lepton flavor
  violation a high energy phenomenon?},''
  \href{http://dx.doi.org/10.1103/PhysRevD.89.051302}{{\em Phys.Rev.D}
  {\bfseries 89} (2014) 051302},
  \href{http://arxiv.org/abs/1308.6789}{{\ttfamily arXiv:1308.6789 [hep-ph]}}.

\bibitem{ATLAS:2019kpx}
{\bfseries ATLAS} Collaboration, G.~Aad {\em et~al.}, ``{Search for heavy
  neutral leptons in decays of $W$ bosons produced in 13 TeV $pp$ collisions
  using prompt and displaced signatures with the ATLAS detector},''
  \href{http://dx.doi.org/10.1007/JHEP10(2019)265}{{\em JHEP} {\bfseries 10}
  (2019) 265}, \href{http://arxiv.org/abs/1905.09787}{{\ttfamily
  arXiv:1905.09787 [hep-ex]}}.

\bibitem{CMS:2018iaf}
{\bfseries CMS} Collaboration, A.~M. Sirunyan {\em et~al.}, ``{Search for heavy
  neutral leptons in events with three charged leptons in proton-proton
  collisions at $\sqrt{s} =$ 13 TeV},''
  \href{http://dx.doi.org/10.1103/PhysRevLett.120.221801}{{\em Phys.Rev.Lett.}
  {\bfseries 120} (2018) 221801},
  \href{http://arxiv.org/abs/1802.02965}{{\ttfamily arXiv:1802.02965
  [hep-ex]}}.

\bibitem{CMS:2022nty}
{\bfseries CMS} Collaboration, A.~Tumasyan {\em et~al.}, ``{Inclusive
  nonresonant multilepton probes of new phenomena at s=13{\,}{\,}TeV},''
  \href{http://dx.doi.org/10.1103/PhysRevD.105.112007}{{\em Phys.Rev.D}
  {\bfseries 105} no.~11, (2022) 112007},
  \href{http://arxiv.org/abs/2202.08676}{{\ttfamily arXiv:2202.08676
  [hep-ex]}}.

\bibitem{Drewes:2019fou}
M.~Drewes and J.~Hajer, ``{Heavy Neutrinos in displaced vertex searches at the
  LHC and HL-LHC},'' \href{http://dx.doi.org/10.1007/JHEP02(2020)070}{{\em
  JHEP} {\bfseries 02} (2020) 070},
  \href{http://arxiv.org/abs/1903.06100}{{\ttfamily arXiv:1903.06100
  [hep-ph]}}.

\bibitem{Abdullahi:2022jlv}
A.~M. Abdullahi {\em et~al.}, ``{The Present and Future Status of Heavy Neutral
  Leptons},'' in {\em {2022 Snowmass Summer Study}}.
\newblock 3, 2022.
\newblock \href{http://arxiv.org/abs/2203.08039}{{\ttfamily arXiv:2203.08039
  [hep-ph]}}.

\bibitem{Papoulias:2018uzy}
D.~K. Papoulias, R.~Sahu, T.~S. Kosmas, V.~K.~B. Kota, and B.~Nayak, ``{Novel
  neutrino-floor and dark matter searches with deformed shell model
  calculations},'' \href{http://dx.doi.org/10.1155/2018/6031362}{{\em Adv. High
  Energy Phys.} {\bfseries 2018} (2018) 6031362},
  \href{http://arxiv.org/abs/1804.11319}{{\ttfamily arXiv:1804.11319
  [hep-ph]}}.

\bibitem{Drukier:1984vhf}
A.~Drukier and L.~Stodolsky, ``{Principles and Applications of a Neutral
  Current Detector for Neutrino Physics and Astronomy},''
  \href{http://dx.doi.org/10.1103/PhysRevD.30.2295}{{\em Phys. Rev. D}
  {\bfseries 30} (1984) 2295}.

\bibitem{Barranco:2005yy}
J.~Barranco, O.~G. Miranda, and T.~I. Rashba, ``{Probing new physics with
  coherent neutrino scattering off nuclei},''
  \href{http://dx.doi.org/10.1088/1126-6708/2005/12/021}{{\em JHEP} {\bfseries
  12} (2005) 021}, \href{http://arxiv.org/abs/hep-ph/0508299}{{\ttfamily
  arXiv:hep-ph/0508299}}.

\bibitem{Papoulias:2019xaw}
D.~K. Papoulias, T.~S. Kosmas, and Y.~Kuno, ``{Recent probes of standard and
  non-standard neutrino physics with nuclei},''
  \href{http://dx.doi.org/10.3389/fphy.2019.00191}{{\em Front. in Phys.}
  {\bfseries 7} (2019) 191}, \href{http://arxiv.org/abs/1911.00916}{{\ttfamily
  arXiv:1911.00916 [hep-ph]}}.

\bibitem{Papoulias:2015vxa}
D.~K. Papoulias and T.~S. Kosmas, ``{Standard and Nonstandard Neutrino-Nucleus
  Reactions Cross Sections and Event Rates to Neutrino Detection
  Experiments},'' \href{http://dx.doi.org/10.1155/2015/763648}{{\em Adv. High
  Energy Phys.} {\bfseries 2015} (2015) 763648},
  \href{http://arxiv.org/abs/1502.02928}{{\ttfamily arXiv:1502.02928
  [nucl-th]}}.

\bibitem{Weinberg:1967tq}
S.~Weinberg, ``{A Model of Leptons},''
  \href{http://dx.doi.org/10.1103/PhysRevLett.19.1264}{{\em Phys. Rev. Lett.}
  {\bfseries 19} (1967) 1264--1266}.

\bibitem{10.1093/ptep/ptaa104}
{\bfseries Particle Data Group} Collaboration, R.~L. Workman, ``{Review of
  Particle Physics},'' \href{http://dx.doi.org/10.1093/ptep/ptac097}{{\em PTEP}
  {\bfseries 2022} (2022) 083C01}.

\bibitem{Erler:2019hds}
J.~Erler and M.~Schott, ``{Electroweak Precision Tests of the Standard Model
  after the Discovery of the Higgs Boson},''
  \href{http://dx.doi.org/10.1016/j.ppnp.2019.02.007}{{\em Prog. Part. Nucl.
  Phys.} {\bfseries 106} (2019) 68--119},
  \href{http://arxiv.org/abs/1902.05142}{{\ttfamily arXiv:1902.05142
  [hep-ph]}}.

\bibitem{Cadeddu:2021ijh}
M.~Cadeddu, N.~Cargioli, F.~Dordei, C.~Giunti, Y.~F. Li, E.~Picciau, C.~A.
  Ternes, and Y.~Y. Zhang, ``{New insights into nuclear physics and weak mixing
  angle using electroweak probes},''
  \href{http://dx.doi.org/10.1103/PhysRevC.104.065502}{{\em Phys. Rev. C}
  {\bfseries 104} no.~6, (2021) 065502},
  \href{http://arxiv.org/abs/2102.06153}{{\ttfamily arXiv:2102.06153
  [hep-ph]}}.

\bibitem{Wood:1997zq}
C.~S. Wood, S.~C. Bennett, D.~Cho, B.~P. Masterson, J.~L. Roberts, C.~E.
  Tanner, and C.~E. Wieman, ``{Measurement of parity nonconservation and an
  anapole moment in cesium},''
  \href{http://dx.doi.org/10.1126/science.275.5307.1759}{{\em Science}
  {\bfseries 275} (1997) 1759--1763}.

\bibitem{Guena:2004sq}
J.~Guena, M.~Lintz, and M.~A. Bouchiat, ``{Measurement of the parity violating
  6S-7S transition amplitude in cesium achieved within 2 x 10(-13) atomic-unit
  accuracy by stimulated-emission detection},''
  \href{http://dx.doi.org/10.1103/PhysRevA.71.042108}{{\em Phys. Rev. A}
  {\bfseries 71} (2005) 042108},
  \href{http://arxiv.org/abs/physics/0412017}{{\ttfamily
  arXiv:physics/0412017}}.

\bibitem{Schechter:1980gr}
J.~Schechter and J.~W.~F. Valle, ``{Neutrino Masses in SU(2) x U(1)
  Theories},'' \href{http://dx.doi.org/10.1103/PhysRevD.22.2227}{{\em Phys.
  Rev. D} {\bfseries 22} (1980) 2227}.

\bibitem{Schechter:1981cv}
J.~Schechter and J.~W.~F. Valle, ``{Neutrino Decay and Spontaneous Violation of
  Lepton Number},'' \href{http://dx.doi.org/10.1103/PhysRevD.25.774}{{\em
  Phys.Rev.D} {\bfseries 25} (1982) 774}.

\bibitem{Escrihuela:2015wra}
F.~J. Escrihuela, D.~V. Forero, O.~G. Miranda, M.~Tortola, and J.~W.~F. Valle,
  ``{On the description of nonunitary neutrino mixing},''
  \href{http://dx.doi.org/10.1103/PhysRevD.92.053009}{{\em Phys. Rev. D}
  {\bfseries 92} no.~5, (2015) 053009},
  \href{http://arxiv.org/abs/1503.08879}{{\ttfamily arXiv:1503.08879
  [hep-ph]}}. [Erratum: Phys.Rev.D 93, 119905 (2016)].

\bibitem{Miranda:2020syh}
O.~G. Miranda, D.~K. Papoulias, O.~Sanders, M.~T\'ortola, and J.~W.~F. Valle,
  ``{Future CEvNS experiments as probes of lepton unitarity and light-sterile
  neutrinos},'' \href{http://dx.doi.org/10.1103/PhysRevD.102.113014}{{\em Phys.
  Rev. D} {\bfseries 102} (2020) 113014},
  \href{http://arxiv.org/abs/2008.02759}{{\ttfamily arXiv:2008.02759
  [hep-ph]}}.

\bibitem{Alfonso-Pita:2022eli}
E.~Alfonso-Pita, L.~J. Flores, E.~Peinado, and E.~V\'azquez-J\'auregui, ``{New
  physics searches in a low threshold scintillating argon bubble chamber
  measuring coherent elastic neutrino-nucleus scattering in reactors},''
  \href{http://dx.doi.org/10.1103/PhysRevD.105.113005}{{\em Phys. Rev. D}
  {\bfseries 105} no.~11, (2022) 113005},
  \href{http://arxiv.org/abs/2203.05982}{{\ttfamily arXiv:2203.05982
  [hep-ph]}}.

\bibitem{Wolfenstein:1977ue}
L.~Wolfenstein, ``{Neutrino Oscillations in Matter},''
  \href{http://dx.doi.org/10.1103/PhysRevD.17.2369}{{\em Phys.Rev.D} {\bfseries
  17} (1978) 2369--2374}.

\bibitem{Valle:1987gv}
J.~W.~F. Valle, ``{Resonant Oscillations of Massless Neutrinos in Matter},''
  \href{http://dx.doi.org/10.1016/0370-2693(87)90947-6}{{\em Phys.Lett.B}
  {\bfseries 199} (1987) 432--436}.

\bibitem{Lee:1956qn}
T.~D. Lee and C.-N. Yang, ``{Question of Parity Conservation in Weak
  Interactions},'' \href{http://dx.doi.org/10.1103/PhysRev.104.254}{{\em Phys.
  Rev.} {\bfseries 104} (1956) 254--258}.

\bibitem{Ohlsson:2012kf}
T.~Ohlsson, ``{Status of non-standard neutrino interactions},''
  \href{http://dx.doi.org/10.1088/0034-4885/76/4/044201}{{\em Rept. Prog.
  Phys.} {\bfseries 76} (2013) 044201},
  \href{http://arxiv.org/abs/1209.2710}{{\ttfamily arXiv:1209.2710 [hep-ph]}}.

\bibitem{Farzan:2017xzy}
Y.~Farzan and M.~Tortola, ``{Neutrino oscillations and Non-Standard
  Interactions},'' \href{http://dx.doi.org/10.3389/fphy.2018.00010}{{\em Front.
  in Phys.} {\bfseries 6} (2018) 10},
  \href{http://arxiv.org/abs/1710.09360}{{\ttfamily arXiv:1710.09360
  [hep-ph]}}.

\bibitem{Billard:2018jnl}
J.~Billard, J.~Johnston, and B.~J. Kavanagh, ``{Prospects for exploring New
  Physics in Coherent Elastic Neutrino-Nucleus Scattering},''
  \href{http://dx.doi.org/10.1088/1475-7516/2018/11/016}{{\em JCAP} {\bfseries
  11} (2018) 016}, \href{http://arxiv.org/abs/1805.01798}{{\ttfamily
  arXiv:1805.01798 [hep-ph]}}.

\bibitem{Papoulias:2017qdn}
D.~K. Papoulias and T.~S. Kosmas, ``{COHERENT constraints to conventional and
  exotic neutrino physics},''
  \href{http://dx.doi.org/10.1103/PhysRevD.97.033003}{{\em Phys. Rev. D}
  {\bfseries 97} no.~3, (2018) 033003},
  \href{http://arxiv.org/abs/1711.09773}{{\ttfamily arXiv:1711.09773
  [hep-ph]}}.

\bibitem{Papoulias:2019txv}
D.~K. Papoulias, ``{COHERENT constraints after the COHERENT-2020 quenching
  factor measurement},''
  \href{http://dx.doi.org/10.1103/PhysRevD.102.113004}{{\em Phys.Rev.}
  {\bfseries D102} (2020) 113004},
  \href{http://arxiv.org/abs/1907.11644}{{\ttfamily arXiv:1907.11644
  [hep-ph]}}.

\bibitem{Miranda:2020tif}
O.~Miranda, D.~Papoulias, G.~Sanchez~Garcia, O.~Sanders, M.~T{\'o}rtola, and
  J.~W.~F. Valle, ``{Implications of the first detection of coherent elastic
  neutrino-nucleus scattering (CEvNS) with Liquid Argon},''
  \href{http://dx.doi.org/10.1007/JHEP05(2020)130}{{\em JHEP} {\bfseries 2005}
  (2020) 130}, \href{http://arxiv.org/abs/2003.12050}{{\ttfamily
  arXiv:2003.12050 [hep-ph]}}.

\bibitem{Giunti:2019xpr}
C.~Giunti, ``{General COHERENT constraints on neutrino nonstandard
  interactions},'' \href{http://dx.doi.org/10.1103/PhysRevD.101.035039}{{\em
  Phys.Rev.} {\bfseries D101} (2020) 035039},
  \href{http://arxiv.org/abs/1909.00466}{{\ttfamily arXiv:1909.00466
  [hep-ph]}}.

\bibitem{Khan:2021wzy}
A.~N. Khan, D.~W. McKay, and W.~Rodejohann, ``{CP-violating and charged current
  neutrino nonstandard interactions in CE{\ensuremath{\nu}}NS},''
  \href{http://dx.doi.org/10.1103/PhysRevD.104.015019}{{\em Phys.Rev.D}
  {\bfseries 104} (2021) 015019},
  \href{http://arxiv.org/abs/2104.00425}{{\ttfamily arXiv:2104.00425
  [hep-ph]}}.

\bibitem{Dutta:2020che}
B.~Dutta, R.~F. Lang, S.~Liao, S.~Sinha, L.~Strigari, and A.~Thompson, ``{A
  global analysis strategy to resolve neutrino NSI degeneracies with scattering
  and oscillation data},''
  \href{http://dx.doi.org/10.1007/JHEP09(2020)106}{{\em JHEP} {\bfseries 2009}
  (2020) 106}, \href{http://arxiv.org/abs/2002.03066}{{\ttfamily
  arXiv:2002.03066 [hep-ph]}}.

\bibitem{Canas:2019fjw}
B.~Canas, E.~Garces, O.~Miranda, A.~Parada, and G.~Sanchez~Garcia, ``{Interplay
  between nonstandard and nuclear constraints in coherent elastic
  neutrino-nucleus scattering experiments},''
  \href{http://dx.doi.org/10.1103/PhysRevD.101.035012}{{\em Phys.Rev.}
  {\bfseries D101} (2020) 035012},
  \href{http://arxiv.org/abs/1911.09831}{{\ttfamily arXiv:1911.09831
  [hep-ph]}}.

\bibitem{Coloma:2019mbs}
P.~Coloma, I.~Esteban, M.~Gonzalez-Garcia, and M.~Maltoni, ``{Improved global
  fit to Non-Standard neutrino Interactions using COHERENT energy and timing
  data},'' \href{http://dx.doi.org/10.1007/JHEP02(2020)023}{{\em JHEP}
  {\bfseries 2002} (2020) 023},
  \href{http://arxiv.org/abs/1911.09109}{{\ttfamily arXiv:1911.09109
  [hep-ph]}}.

\bibitem{Chatterjee:2022mmu}
S.~S. Chatterjee, S.~Lavignac, O.~Miranda, and G.~Sanchez~Garcia,
  ``{Constraining Non-Standard Interactions with Coherent Elastic
  Neutrino-Nucleus Scattering at the European Spallation Source},''
  \href{http://arxiv.org/abs/2208.11771}{{\ttfamily arXiv:2208.11771
  [hep-ph]}}.

\bibitem{Lindner:2016wff}
M.~Lindner, W.~Rodejohann, and X.-J. Xu, ``{Coherent Neutrino-Nucleus
  Scattering and new Neutrino Interactions},''
  \href{http://dx.doi.org/10.1007/JHEP03(2017)097}{{\em JHEP} {\bfseries 03}
  (2017) 097}, \href{http://arxiv.org/abs/1612.04150}{{\ttfamily
  arXiv:1612.04150 [hep-ph]}}.

\bibitem{Jenkins:2017jig}
E.~E. Jenkins, A.~V. Manohar, and P.~Stoffer, ``{Low-Energy Effective Field
  Theory below the Electroweak Scale: Operators and Matching},''
  \href{http://dx.doi.org/10.1007/JHEP03(2018)016}{{\em JHEP} {\bfseries 03}
  (2018) 016}, \href{http://arxiv.org/abs/1709.04486}{{\ttfamily
  arXiv:1709.04486 [hep-ph]}}.

\bibitem{AristizabalSierra:2018eqm}
D.~Aristizabal~Sierra, V.~De~Romeri, and N.~Rojas, ``{COHERENT analysis of
  neutrino generalized interactions},''
  \href{http://dx.doi.org/10.1103/PhysRevD.98.075018}{{\em Phys. Rev. D}
  {\bfseries 98} (2018) 075018},
  \href{http://arxiv.org/abs/1806.07424}{{\ttfamily arXiv:1806.07424
  [hep-ph]}}.

\bibitem{Dent:2015zpa}
J.~B. Dent, L.~M. Krauss, J.~L. Newstead, and S.~Sabharwal, ``{General analysis
  of direct dark matter detection: From microphysics to observational
  signatures},'' \href{http://dx.doi.org/10.1103/PhysRevD.92.063515}{{\em Phys.
  Rev. D} {\bfseries 92} no.~6, (2015) 063515},
  \href{http://arxiv.org/abs/1505.03117}{{\ttfamily arXiv:1505.03117
  [hep-ph]}}.

\bibitem{Cirelli:2013ufw}
M.~Cirelli, E.~Del~Nobile, and P.~Panci, ``{Tools for model-independent bounds
  in direct dark matter searches},''
  \href{http://dx.doi.org/10.1088/1475-7516/2013/10/019}{{\em JCAP} {\bfseries
  10} (2013) 019}, \href{http://arxiv.org/abs/1307.5955}{{\ttfamily
  arXiv:1307.5955 [hep-ph]}}.

\bibitem{AristizabalSierra:2019zmy}
D.~Aristizabal~Sierra, J.~Liao, and D.~Marfatia, ``{Impact of form factor
  uncertainties on interpretations of coherent elastic neutrino-nucleus
  scattering data},'' \href{http://dx.doi.org/10.1007/JHEP06(2019)141}{{\em
  JHEP} {\bfseries 06} (2019) 141},
  \href{http://arxiv.org/abs/1902.07398}{{\ttfamily arXiv:1902.07398
  [hep-ph]}}.

\bibitem{Flores:2021kzl}
L.~J. Flores, N.~Nath, and E.~Peinado, ``{CE\ensuremath{\nu}NS as a probe of
  flavored generalized neutrino interactions},''
  \href{http://dx.doi.org/10.1103/PhysRevD.105.055010}{{\em Phys. Rev. D}
  {\bfseries 105} no.~5, (2022) 055010},
  \href{http://arxiv.org/abs/2112.05103}{{\ttfamily arXiv:2112.05103
  [hep-ph]}}.

\bibitem{CONNIE:2019xid}
{\bfseries CONNIE} Collaboration, A.~Aguilar-Arevalo {\em et~al.}, ``{Search
  for light mediators in the low-energy data of the CONNIE reactor neutrino
  experiment},'' \href{http://dx.doi.org/10.1007/JHEP04(2020)054}{{\em JHEP}
  {\bfseries 04} (2020) 054}, \href{http://arxiv.org/abs/1910.04951}{{\ttfamily
  arXiv:1910.04951 [hep-ex]}}.

\bibitem{ANTMAN1966272}
S.~Antman, D.~Landis, and R.~Pehl, ``Measurements of the fano factor and the
  energy per hole-electron pair in germanium,''
  \href{http://dx.doi.org/https://doi.org/10.1016/0029-554X(66)90386-7}{{\em
  Nuclear Instruments and Methods} {\bfseries 40} no.~2, (1966) 272--276}.
  \url{https://www.sciencedirect.com/science/article/pii/0029554X66903867}.

\bibitem{Wei:2016xbw}
W.~Z. Wei, L.~Wang, and D.~M. Mei, ``{Average energy expended per e-h pair for
  germanium-based dark matter experiments},''
  \href{http://dx.doi.org/10.1088/1748-0221/12/04/P04022}{{\em JINST}
  {\bfseries 12} no.~04, (2017) P04022},
  \href{http://arxiv.org/abs/1602.08005}{{\ttfamily arXiv:1602.08005
  [physics.ins-det]}}.

\bibitem{BerglundWieser:2011}
M.~Berglund and M.~E. Wieser, ``Isotopic compositions of the elements 2009
  (iupac technical report),''
  \href{http://dx.doi.org/10.1351/PAC-REP-10-06-02}{{\em Pure and Applied
  Chemistry} {\bfseries 83} no.~2, (2011) 397--410}.

\bibitem{Scholz:2016qos}
B.~J. Scholz, A.~E. Chavarria, J.~I. Collar, P.~Privitera, and A.~E. Robinson,
  ``{Measurement of the low-energy quenching factor in germanium using an
  $^{88}$Y/Be photoneutron source},''
  \href{http://dx.doi.org/10.1103/PhysRevD.94.122003}{{\em Phys. Rev. D}
  {\bfseries 94} no.~12, (2016) 122003},
  \href{http://arxiv.org/abs/1608.03588}{{\ttfamily arXiv:1608.03588
  [physics.ins-det]}}.

\bibitem{Collar:2021fcl}
J.~I. Collar, A.~R.~L. Kavner, and C.~M. Lewis, ``{Germanium response to
  sub-keV nuclear recoils: a multipronged experimental characterization},''
  \href{http://dx.doi.org/10.1103/PhysRevD.103.122003}{{\em Phys. Rev. D}
  {\bfseries 103} no.~12, (2021) 122003},
  \href{http://arxiv.org/abs/2102.10089}{{\ttfamily arXiv:2102.10089
  [nucl-ex]}}.

\bibitem{SchoenfeldSep1998}
E.~Schoenfeld, ``Calculation of fractional electron capture probabilities,''
  {\em Applied Radiation and Isotopes} {\bfseries 49} no.~9-11, (Sep 1998)
  1353--1357. \url{http://inis.iaea.org/search/search.aspx?orig_q=RN:34013588}.
  Nuclear physics and radiation physics.

\bibitem{SuperCDMS:2015eex}
{\bfseries SuperCDMS} Collaboration, R.~Agnese {\em et~al.}, ``{New Results
  from the Search for Low-Mass Weakly Interacting Massive Particles with the
  CDMS Low Ionization Threshold Experiment},''
  \href{http://dx.doi.org/10.1103/PhysRevLett.116.071301}{{\em Phys. Rev.
  Lett.} {\bfseries 116} no.~7, (2016) 071301},
  \href{http://arxiv.org/abs/1509.02448}{{\ttfamily arXiv:1509.02448
  [astro-ph.CO]}}.

\bibitem{Escrihuela:2016ube}
F.~J. Escrihuela {\em et~al.}, ``{Probing CP violation with non-unitary mixing
  in long-baseline neutrino oscillation experiments: DUNE as a case study},''
  \href{http://dx.doi.org/10.1088/1367-2630/aa79ec}{{\em New J. Phys.}
  {\bfseries 19} no.~9, (2017) 093005},
  \href{http://arxiv.org/abs/1612.07377}{{\ttfamily arXiv:1612.07377
  [hep-ph]}}.

\end{thebibliography}\endgroup

\end{document}